\documentclass[prl,twocolumn,amsmath,amssymb,floatfix]{revtex4}

\usepackage{graphicx}
\usepackage{bm}
\usepackage{natbib}
\usepackage{braket}

\newcommand{\be}{\begin{equation}}
\newcommand{\ee}{\end{equation}}
\newcommand{\beq}{\begin{eqnarray}}
\newcommand{\eeq}{\end{eqnarray}}

\usepackage{amsmath}
\usepackage{braket}
\usepackage{graphicx} 
\usepackage{color}

\def\url#1{}

\usepackage{array} 
\usepackage{paralist} 
\usepackage{verbatim} 

\begin{document}

\title{Cooling and Autonomous Feedback in a Bose-Hubbard Chain with Attractive Interactions}


\author{S. Hacohen-Gourgy$^1$}
\email[shayhh@berkeley.edu]{shayhh@berkeley.edu}
\author{V. V. Ramasesh$^1$}
\author{C. De Grandi$^2$}
\author{I. Siddiqi$^1$}
\author{S. M. Girvin$^2$}
\affiliation{$^1$Quantum Nanoelectronics Laboratory, Department of Physics, University of California, Berkeley CA 94720\\
$^2$Departments of Physics and Applied Physics, Yale University, New Haven, Connecticut 06520, USA}

\date{\today}

\begin{abstract}

We engineer a quantum bath that enables entropy and energy exchange with a one-dimensional Bose-Hubbard  lattice
with attractive on-site interactions. We implement this in an array of three superconducting transmon qubits coupled to a single cavity mode; the transmons represent lattice sites and their excitation quanta embody bosonic particles. 
 Our cooling protocol preserves particle number---realizing a canonical ensemble--- and also affords the efficient preparation of dark states which, due to symmetry, cannot be prepared via coherent drives on the cavity. Furthermore, by applying continuous microwave radiation, we also realize autonomous feedback to indefinitely stabilize particular eigenstates of the array.

\end{abstract}

\maketitle


Ordinarily, uncontrolled dissipation destroys quantum coherence, but it is now appreciated that an engineered quantum bath is a valuable resource for quantum computation (state preparation~ \cite{Kraus-Zoller-PhysRevA.78.042307,Verstraete2009,Murch_PRL,Shankar_al_Nature} and system reset\cite{Geerlings_PhysRevLett.110.120501}) and quantum error correction~\cite{EliotArxiv}.  A dynamical bath can induce cooling or heating, and is of great utility in optomechanical ground state preparation~\cite{Teufelgroundstatecooling,Sillanpaa_groundstatecooling2012}, quantum state transfer between various EM modes~ \cite{Abdo-Devoret-PhysRevLett.110.173902,KonradCoherentStateTransfer2010Nature,KonradBidirectionalNaturePhys2014}, amplification~\cite{Bergeal_Devoret_QuantumAmp_Nature2010,KonradAmp,Sillanpaa_mechanical_amplifier}, many-particle quantum simulation~\cite{WeitzPhotonBEC,Marcos-Rabl-NJP2012}, and entanglement generation~ \cite{Tureci_PRA, Tureci_arxiv}.

\begin{figure}[!ht]
\includegraphics[totalheight=0.54\textwidth]{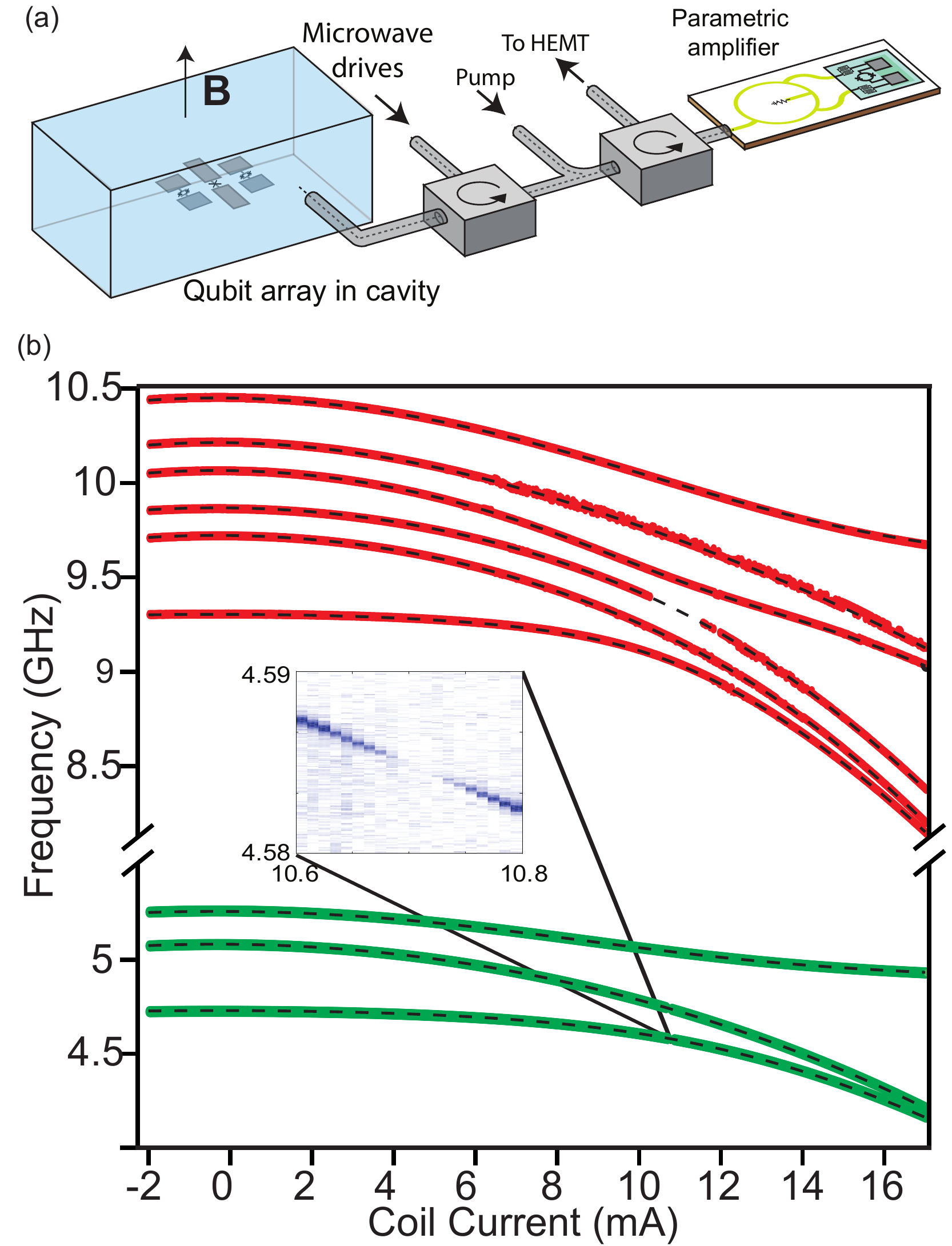}
\caption{(a) A schematic (not to scale) of the experimental setup described in the text.  (b) Spectroscopically measured eigenfrequencies of the one- and two-particle states of the array as a function of current through the external bias coil.  For a given current, the flux through the two SQUIDs in the array differs by 2.5\%; 17 mA roughly corresponds to a quarter of a flux quantum.  Solid lines denote measured frequencies with fits to the 1D Bose-Hubbard Hamiltonian shown as overlaid dashed lines. Red lines correspond to two-particle states; green lines are one-particle states.  The inset shows raw data near the $\ket{E_1}$ frequency, from which the darkness of the $\ket{G}\rightarrow\ket{E_1}$ transition discussed in the text becomes apparent.}
\label{fig:Spec}
\end{figure}

Initially conceived and implemented in trapped ion systems~\cite{Poyatos-Zoller-PhysRevLett.77.4728,Wineland2000}, dissipation engineering has been implemented in solid state systems with two recent experiments on superconducting qubits: autonomously controlling the orientation on the Bloch sphere of a single qubit~\cite{Murch_PRL}, and stabilizing a Bell-state in a two-qubit system~\cite{Shankar_al_Nature}.  In this Letter, we experimentally demonstrate that dissipation engineering can be used to control a novel, complex superconducting system embodying a much larger Hilbert space: the ten lowest-lying energy levels of a coupled, three-transmon array which realizes a one-dimensional, attractive Bose-Hubbard Hamiltonian.  The techniques employed in this work define a path for cooling and stabilizing complex quantum systems to specific target states. 

The Bose-Hubbard Hamiltonian~\cite{Fisher_et_al_PhysRevB.40.546} is a prototypical model used to describe a broad class of quantum matter.  While the repulsive side of the Bose-Hubbard phase diagram has been extensively explored in groundbreaking experiments with ultracold atoms in optical lattices~\cite{Greiner,PRL_1D_SF_mott,RevModPhys_cold_atoms}, the attractive regime has thus far eluded emulation. Our realization of this model here, in perhaps its simplest incarnation, opens the door to experimental verification of as-yet unexplored predictions of attractive Bose-Hubbard dynamics: the existence of self-bound states~\cite{Buonsante_PRA72, Buonsante_PRA82, Jack_PRA71} and the possibility to create large-scale multipartite entanglement~\cite{AttractiveBH}. 

\begin{figure}[!ht]
\includegraphics[totalheight=0.38\textwidth]{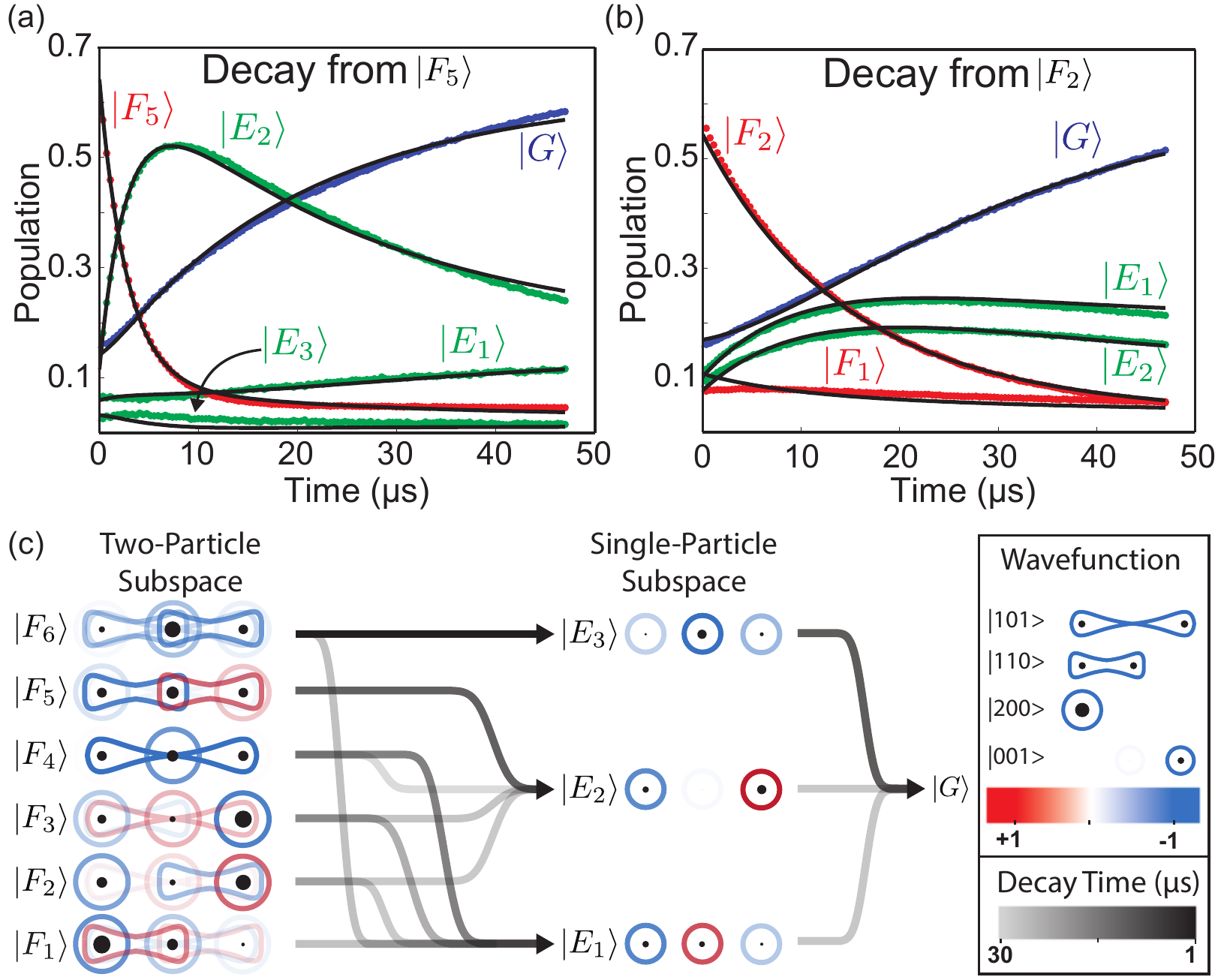}
\caption{(a) and (b) Examples of spontaneous decays from two-particle states to the global ground state via the one-particle subspace.  In (a), $\ket{F_5}$ decays rapidly and almost entirely to $\ket{E_2}$, which then decays to $\ket{G}$, while in (b), $\ket{F_2}$ decays with roughly equal rates to $\ket{E_1}$ and $\ket{E_2}$.  Overlaid black lines are obtained from fitting the decay data for all nine excited states to a single rate-equation model, as described in the supplement~\cite{SM}.  (c) An illustration of the natural decay pathways of the system, from the two-particle subspace on the left, through the one-particle subspace in the middle, to the zero-particle state on the right.  Each black arrow represents a decay channel, with the opacity of the line indicating the rate of the transition.  Darker lines indicate faster rates, as the legend shows. Also shown are representations of the eigenstate wavefunctions in the qubit basis. The black circle radius is proportional to the mean particle number.}
\label{fig:Decays}
\end{figure}

The cooling protocol we develop here, based on Raman scattering processes, facilitates entropy and energy exchange between the qubit array and its bath while preserving the total number of excitation quanta in the array.  Essentially, this amounts to simulation within the canonical ensemble; a similar path to grand-canonical simulation via bath engineering (using photons) has recently been proposed theoretically by Hafezi et al~\cite{Taylor}.  In another related work, a cooling scheme similar to ours has been employed recently to measure the dynamic structure factor of a gas of cold atoms~\cite{Esslinger}.  Additionally, a recent experiment with superconducting qubits has achieved simulation of the unitary Fermi-Hubbard dynamics via discrete gates~\cite{Barends2015}. 

Our system is comprised of an array of three capacitively-coupled transmon qubits~\cite{PhysRevA.76.042319}, each coupled dispersively~\cite{PhysRevLett.95.060501} to a waveguide cavity~\cite{Paik}. The two qubits on the ends of the array utilize a SQUID loop to allow tuning their frequencies via an external magnetic field. Taking into account the full transmon spectrum~\cite{PhysRevA.76.042319}, the system is described by the Hamiltonian $H = H_\mathrm{cav} + H_\mathrm{array} + H_\mathrm{int}$, with
\be\label{H_cav}
H_{\mathrm{cav}} = \hbar\omega_c \left( a^{\dagger}a + 1/2 \right)
\ee

\be\label{H_array}
\begin{array}{rcl}
H_\mathrm{array} &=&  \hbar\sum\limits_{j=1}^3 \left(\omega_{j}b_{j}^{\dagger}b_{j} +\frac{\alpha_{j}}{2}b_{j}^{\dagger}b_{j}^{\dagger} b_{j} b_{j}\right) \\
 &&  +\hbar J\sum\limits_{j=1}^{2} (b_{j+1}^{\dagger}b_{j}+b_{j}^{\dagger}b_{j+1}) \\
 &&  +\hbar J_{13} (b_{1}^{\dagger}b_{3}+b_{3}^{\dagger}b_{1})
\end{array}
\ee

\be\label{H_int}
H_\mathrm{int} =  \hbar\sum\limits_{j=1}^3 g_{j}(b_{j}a^{\dagger}+b_{j}^{\dagger}a),
\ee
where $a^{\dagger}$ and $b_{j}^{\dagger}$ are creation operators for cavity photons and excitation quanta of the $j^{th}$ qubit in the array, respectively.  $H_\mathrm{cav}$ is the Hamiltonian for the 7.116 GHz waveguide cavity, which couples dispersively to each qubit in the transmon array with strength $g_j$, as described by $H_\mathrm{int}$.  The array itself is described by $H_\mathrm{array}$: each transmon is a weakly anharmonic oscillator with $|0\rangle\rightarrow|1\rangle$ transition frequency $\omega_j$ and (negative) anharmonicity $\alpha_j$.  The three-qubit system is effectively an array of lattice sites on which particles---the transmon excitation quanta---can hop with nearest-neighbor tunneling strength $\hbar J$, and next-nearest-neighbor tunneling $\hbar J_{13}$ (set primarily by the capactive coupling~\cite{PhysRevLett.110.030601}), with $J_{13} \ll J$.  In this language the negative anharmonicity of the transmon gives rise to an attractive pairwise interaction between these particles, since distributing a pair of excitations among two identical transmons takes an energy $\alpha$ more than lumping the quanta together on the same qubit.  It is this combination of tunneling and on-site interaction which realizes the canonical 1D Bose-Hubbard Hamiltonian~\cite{Fisher_et_al_PhysRevB.40.546}. 

\begin{figure*} [ht]
\includegraphics[totalheight=0.45\textwidth]{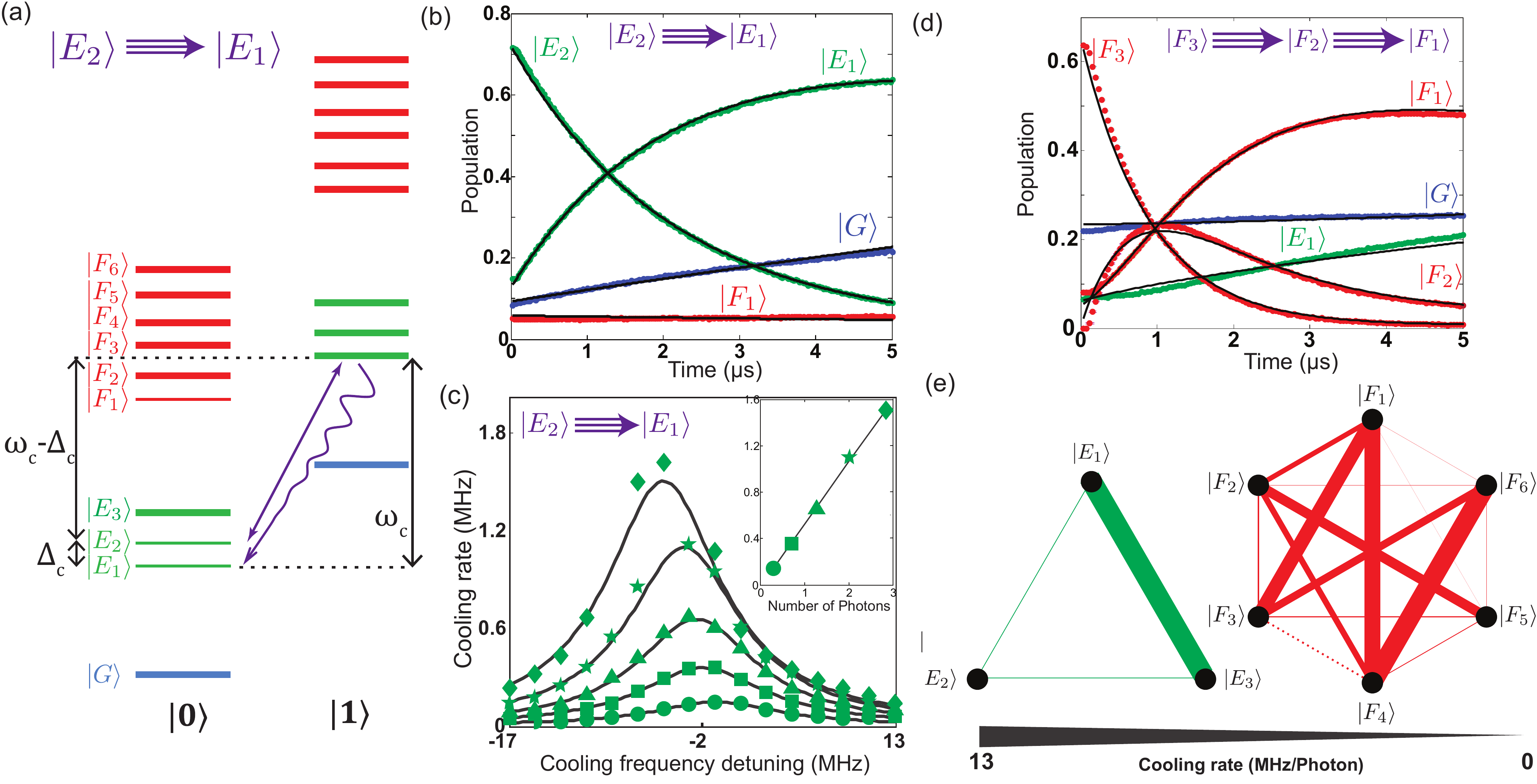}
\caption{(a) An approximate representation of the eigenstates of the array-cavity system; states in the left-hand (right-hand) column contain zero (one) cavity photons.  In this picture, our cooling process can be understood qualitatively as follows (taking $\ket{E_2}\rightarrow\ket{E_1}$) as an example:  the cooling drive at frequency $\omega_{\mathrm{c}} - \Delta_c$ induces a transition from $\ket{E_2}\ket{0}$ to $\ket{E_1}\ket{1}$, where the second ket indicates the cavity photon number.  The cavity state $\ket{1}$ decays via photon emission, leaving the system in the state $\ket{E_1}\ket{0}$, as desired. (b) Example of cooling from  $\ket{E_2}\rightarrow\ket{E_1}$ at an incident power corresponding to 1.3 drive photons in the cavity. (c) The cooling rate versus drive frequency is Lorentzian, centered around $\omega_{\mathrm{c}} - (E_2 - E_1)$, with linewidth roughly $\kappa$.  As the drive power is increased, the Lorentzian peak shifts due to a Stark shift of the relevant transition frequency. The inset shows that in the regime where the cooling rate is much lower than $\kappa$, the rate scales linearly with incident power. (d) Example of cascaded cooling, where the system is cooled from $|F_3\rangle$ to $|F_1\rangle$ via the intermediate state $|F_2\rangle$.  (e) Connected graphs representing the measured cooling rates per photon in the linear regime for the one- (left) and two- (right) particle subspaces, with the width of the line indicating the rate of the corresponding transition. The dispersive shifts for the $\ket{F_3}$ and $\ket{F_4}$ states are almost identical, so they cannot be distinguished by our measurement. We thus do not measure cooling from $\ket{F_4}$ to $\ket{F_3}$.}
\label{fig:Cooling}
\end{figure*}

\begin{figure}
\includegraphics[totalheight=0.35\textwidth]{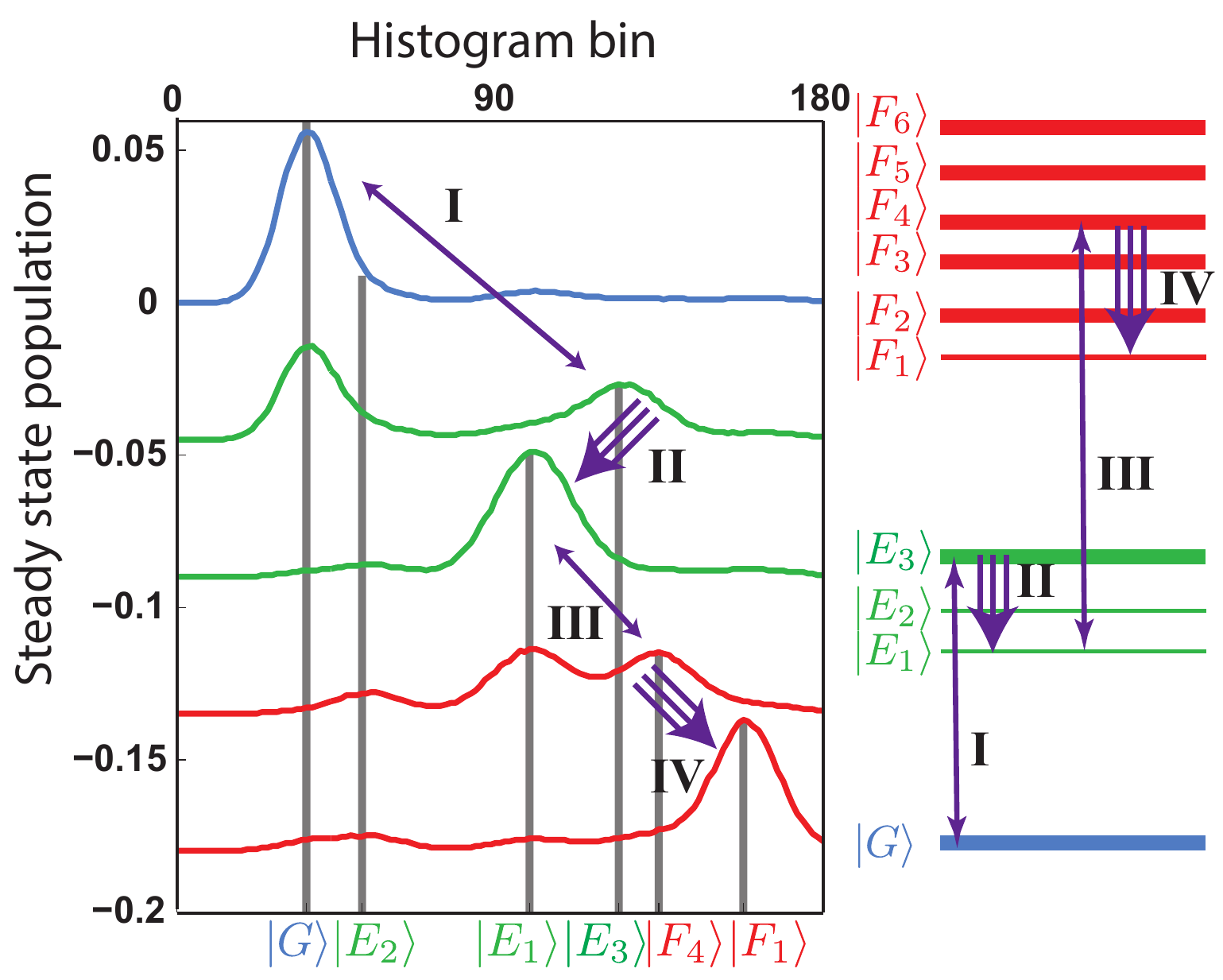}
\caption{Steady state population at the different stages of the persistent stabilization scheme of the two particle ground state $\ket{F_1}$. The top trace shows the thermal equilibrium population, where 78\% of the population is in $\ket{G}$.  In each subsequent trace a further drive is added: first a coherent drive $\ket{G}\rightarrow\ket{E_3}$ (I), then cooling from $\ket{E_3}\rightarrow\ket{E_1}$ (II), coherent drive from $\ket{E_1}\rightarrow\ket{F_4}$ (III), and finally a cooling drive from $\ket{F_4}\rightarrow\ket{F_1}$ (IV).  At the end, 67\% of the population is in the desired state, $\ket{F_1}$.  $\ket{F_4}$ contains the bulk, 13\%, of the residual population.  Single-particle ($\ket{E_1}$) stabilization achieves a population of 80\% (data not shown).}
\label{fig:Stab}
\end{figure}

Since the Bose-Hubbard Hamiltonian conserves total particle number, eigenstates of the three-qubit array can be grouped into manifolds characterized by this quantum number.  In our experiment, we work with the zero, one, and two-particle manifolds, comprising respectively one, three, and six states.  We denote the zero-particle state by $|G\rangle$, the single-particle states by $\{|E_i\rangle \text{,~} i\in [1,3]\}$, and the two-particle states by $\{|F_j\rangle \text{,~} j\in [1,6]\}$, with increasing subscript value indicating higher-energy states.  Due to the dispersive interaction between qubits and the cavity, a reflected microwave signal near the cavity resonance frequency acquires a phase shift dependent on the state of the array.  Amplification of the readout signal via a Josephson parametric amplifier~\cite{PhysRevB.83.134501} and subsequent higher-temperature electronics allows for measurement of the signal phase and hence the array state.

Using this dispersive readout, we first characterize the system by spectroscopically probing its energy levels as the edge qubit frequencies are tuned down via external magnetic flux.  As shown in Fig.~\ref{fig:Spec}, our measurement of the system's one- and two-particle energy states agrees well with predictions based on the attractive 1D Bose-Hubbard Hamiltonian. The extracted parameters are: $\omega_1/2 \pi$ = 5.074 GHz, $\omega_2/2 \pi$ = 4.892 GHz, $\omega_3/2 \pi$ = 5.165 GHz, $J/2 \pi$ = 0.177 GHz and $J_{13}/2 \pi$ = 0.026 GHz, all to within $\pm~$0.003 GHz.  The cavity-qubit coupling $g_1/2 \pi$ = 149 $\pm$ 7  MHz, $g_2/2 \pi$ = 264  $\pm$ 7 MHz, $g_3/2 \pi$ = 155  $\pm$ 7 MHz and qubit anharmonicities $\alpha_{1,3}/2 \pi$ = -214  $\pm$ 1 MHz, $\alpha_{2}/2 \pi$ = -240  $\pm$ 1 MHz, were calibrated independently. While at zero applied flux the qubits are relatively well-separated in frequency, as the edge qubits are tuned down towards the middle qubit, avoided crossings become apparent, showcasing the coupling between the qubits. Our system lies in the parameter regime where the competing tunneling ($J/2 \pi \sim 180~\mathrm{MHz}$) and on-site interactions ($\alpha/2 \pi \sim -220~\mathrm{MHz}$) have nearly equal strength.

At an external field corresponding to 10 mA in the bias coil, the individual qubit frequencies approximately coincide---the detunings between neighboring qubits, 45 and 115 MHz, are lower than the tunneling rate $J$.  At this bias we characterize---in the absence of engineered dissipation---the particle-loss dynamics resulting from coupling to the dissipative environment comprised of both the leaky cavity ($\kappa/2 \pi = 10 \mathrm{MHz}$) and microscopic material imperfections.  A coherent microwave pulse initializes the system to the desired $|E_i\rangle$ or $|F_j\rangle$ state, after which the resulting  populations are measured as a function of time.  As expected, these decay dynamics fit well to a model which only includes single-particle loss events: direct transitions from the $\ket{F_i}$ states to $\ket{G}$ are suppressed, as recently observed with a single transmon qubit~\cite{PhysRevLett.114.010501}.  Examples of such decays are shown in Fig.~\ref{fig:Decays}a-b, while Fig.~\ref{fig:Decays}c shows the full map of decay rates within these subspaces.

A striking feature of the natural decay dynamics is the discrepancy between decay times of different states in the same manifold. For example, $|E_1\rangle$ and $|E_2\rangle$ live for $\sim 30 \mu \text{s}$, while $|E_3\rangle$ decays much more quickly, in $\sim 3 \mu \text{s}$.  This is due to the substantially different dipole transition matrix elements that each single-particle state $\ket{E_i}$ exhibits with respect to the final state $\ket{G}$.  A related consequence of these dipole moments is shown in the inset of Fig.~\ref{fig:Spec}, where at 10.71 mA, $|E_1\rangle$ goes fully dark, i.e. becomes impossible to excite via a coherent microwave pulse.  Under only Purcell decay---in the absence of material losses in the system---such a dark state would live indefinitely, making it attractive for shelving an excitation if it could be readily prepared.  We will return to the preparation of these dark states as one application of our bath engineering protocol, which we now discuss.

By altering the quantum noise spectrum of the bath interacting with the Bose-Hubbard chain, we can modify the system decay dynamics. In our cooling protocol, this alteration takes the form of an additional microwave drive incident on the cavity, red-detuned from the cavity resonance by an amount $\Delta_c$, as illustrated in Fig.~\ref{fig:Cooling}a for the $\ket{E_2}$ to $\ket{E_1}$ transition. This induces quantum photon shot noise whose spectral density peaks at frequency $|\Delta_c|$~\cite{SteveRMP}. When $\hbar\Delta_c$ matches the energy difference between the array's initial state $|i\rangle$ and a lower eigenstate $|f\rangle$, the emission of a photon at the cavity's resonance frequency mediates a so-called cooling transition from $\ket{i}$ to $\ket{f}$.  The energy gained from the cooling transition augments the incident photon energy to allow emission on resonance.  Similarly, a blue-detuned drive induces heating transitions to array states of higher energy. The rate for these processes will depend on both the Raman-transition matrix element between the initial and final array states and on the incident photon flux.  The transition rate increases with photon flux up to a value of $\sim\kappa$, saturating there as the dissipative process requires the emission of a photon by the cavity.  Since $\kappa$ is much larger than most of the natural decay rates, this technique is well suited to driving otherwise inaccessible particle-number-conserving transitions in our system. More details on the cooling protocol can be found in \cite{SM}.

To characterize the cooling dynamics, we initialize the system into an $\ket{E_i}$ or $\ket{F_j}$ eigenstate and subsequently apply a cooling drive for a variable time and measure the state of the array.  Cooling rates are extracted via a fit to a model similar to that used for the natural decays, with additional parameters to capture the induced intramanifold transitions. Because the cavity's density of states exhibits a Lorentzian profile with width $\kappa$, so will the transition rate as a function of cooling drive frequency, as shown for the $\ket{E_2}$ to $\ket{E_1}$ transition in Fig.~\ref{fig:Cooling}c.  

For incident powers which cool at a rate $\Gamma_{i\rightarrow j}$ much lower than $\kappa$, a Fermi Golden Rule calculation~\cite{SM} shows that
\begin{equation}
\Gamma_{i\rightarrow f} \propto P_{\rm in} |M_{if}|^2 \frac{\kappa}{(\omega_i - \omega_f + \Delta_c)^2 + ({\kappa}/{2})^2}~,
\end{equation}
where $P_{\rm in}$ is the incident cooling drive power and $M_{if}$ the matrix element connecting the states $|i\rangle$, $|f\rangle$ of the cooling operator~\cite{SM} which describes the effect of the dissipative bath; in other words $M_{if}$ quantifies the coupling between the states $|i\rangle$ and $|f\rangle$ indirectly via the cross-Kerr terms that couple the qubits with cavity. 
 The predicted linear scaling of the peak $\Gamma$ with $P_{\rm in}$ is shown in the inset of Fig.~\ref{fig:Cooling}c for the $\ket{E_2}\rightarrow\ket{E_1}$ transition.  $|M_{ij}|$ provides a measure of the efficacy of each transition; we map out this value for each pair of eigenstates, showing the results in Fig.~\ref{fig:Cooling}e.

In most cases, applying a drive whose frequency is targeted to cool $|i\rangle$ to $|f\rangle$  has no effect on the decay dynamics of the other states, as most cooling drive frequencies are spaced apart by more than several $\kappa$.  However, when multiple cooling frequencies are separated by less than $\kappa$, a single drive can give rise to a so-called \emph{cascaded cooling} effect, whereby multiple cooling transitions happen in sequence. In our system, for example, the $\ket{F_3} \rightarrow \ket{F_2}$ and the $\ket{F_2} \rightarrow \ket{F_1}$ transitions are separated by only 17 MHz, so a single tone can cause the system to cascade from $|F_3\rangle$ to $|F_1\rangle$ via the intermediate state $|F_2\rangle$, as shown in Fig.~\ref{fig:Cooling}d.  In the specific example shown, since the cooling matrix element between $\ket{F_3}$ and $\ket{F_2}$ is substantially lower than that of $\ket{F_2}$ and $\ket{F_1}$ ($|M_{21}|\sim5|M_{32}|$), we cooled with the drive frequency tuned to the $\ket{F_3}\rightarrow\ket{F_2}$ transition; this achieved approximately equal cooling rates from $\ket{F_3}$ to $\ket{F_2}$ and $\ket{F_2}$ to $\ket{F_1}$.  Cascaded cooling sequences could be useful in larger many-qubit systems with manifolds containing several closely-spaced eigenstates (see \cite{SM} for details).

The transitions $\ket{G}\rightarrow\ket{E_1}$, $\ket{G}\rightarrow\ket{E_2}$, and $\ket{E_i}\rightarrow\ket{F_1}$ do not interact strongly with the electromagnetic environment of the cavity on account of the symmetry of the states; this decoupling is responsible for their relatively long lifetimes.  Correspondingly, however, it is difficult to coherently initialize these states via direct transitions, but our cooling scheme affords their efficient preparation.  To illustrate this, consider the $\ket{G}\rightarrow\ket{E_1}$ transition, which as shown in Fig.~\ref{fig:Spec}, is at its darkest at a flux bias of 10.71 mA.  At this bias point, we use the Raman cooling protocol to prepare $\ket{E_1}$ indirectly via the $\ket{E_3}\rightarrow\ket{E_1}$ cooling transition (data not shown).  Further, by combining coherent drives with Raman cooling, we stabilize $\ket{E_1}$ indefinitely against particle loss.  As the first part of Fig.~\ref{fig:Stab} illustrates, this is done by coherently driving the $|G\rangle$ to $|E_3\rangle$ transition with a Rabi frequency of 7 MHz while applying a drive to cool the $|E_3\rangle\rightarrow |E_1\rangle$ transition at a rate of 3 MHz.  We next use $\ket{E_1}$ as a stepping stone to stabilize the two-particle ground state $\ket{F_1}$, as shown in the lower part of Fig.~\ref{fig:Stab}.  To accomplish this we add two additional drives, an extra coherent drive from $|E_1\rangle\rightarrow |F_4\rangle$ with a Rabi frequency of 7 MHz, and a cooling drive from $|F_4\rangle \rightarrow |F_1\rangle$ with a rate of approximately 3 MHz.  Observed fidelities, while inline with a rate matrix calculation, are primarily limited by spurious thermal population of dark states, which can be reduced by additional cooling tones.  This initialization and maintenance of the array in the ground state of a specific particle-number manifold will be a valuable resource for a hardware simulator.

In conclusion, we have realized a three-qubit transmon array and spectroscopically verified that it obeys an attractive 1D Bose-Hubbard Hamiltonian up to the ten lowest-lying eigenstates, highlighting the use of circuit QED in simulating otherwise challenging quantum systems.  Our developed cooling and stabilization protocols, based on quantum bath engineering--a well-studied phenomenon in quantum optics--afford effective control over this solid state system. The capabilities demonstrated here--engineering decay dynamics and stabilizing particular eigenstates--show that dissipation engineering can be a valuable tool as superconducting circuits scale up in complexity to complement simulators based on cold atoms and trapped ions.      

We thank F. Marquardt for motivating this line of research and acknowledge helpful discussions with U. Vool. VVR acknowledges funding via a NSF graduate student fellowship. Financial support for ongoing quantum circuit development was provided by AFOSR FA9550-12-1-0378 and the current experiment was conducted under a DOE Laboratory Directed Research and Development grant. Funding for theory support is acknowledged from NSF DMR-1301798 and ARO W911NF1410011.

\bibliography{cooling_v2}

\pagebreak
\clearpage
\widetext
\begin{center}
\textbf{\large Supplemental Materials: Cooling and Autonomous Feedback in a Bose-Hubbard chain with Attractive Interactions}
\end{center}
\setcounter{equation}{0}
\setcounter{figure}{0}
\setcounter{table}{0}
\setcounter{page}{1}
\makeatletter
\renewcommand{\theequation}{S\arabic{equation}}
\renewcommand{\thefigure}{S\arabic{figure}}
\renewcommand{\bibnumfmt}[1]{[S#1]}
\renewcommand{\citenumfont}[1]{S#1}

\section{Theory of cooling in a transmon array coupled to a cavity}

Here we describe the theory behind the bath-engineering protocol implemented in the experiment.  We first consider the general case of an array of $L$ transmon qubits, each capacitively coupled to its nearest neighbors.  After showing how, in such an array, an engineered bath can induce transitions between array eigenstates which conserve excitation number, we specialize to the case of a three-site array ($L=3$) used in our experiment.  As we will show, theoretical predictions are in qualitative agreement with experimental observations, with the single-mode approximation of the 3D microwave cavity likely contributing a large part of the discrepancy.

\subsection{General L-site array}



A general array of $L$ capacitively-coupled transmon qubits ($j=1, \dots L$) is described by the Hamiltonian
\be\label{H_array}
H_{\rm array}=\hbar \sum_{j=1}^L \left(\omega_j b_j^\dagger b_j+\frac{\alpha_j}{2}  b_j^\dagger b_j^\dagger b_j b_j \right)+\hbar J\sum_{j=1}^{L-1}(b_{j+1}^\dagger b_j+b_j^\dagger b_{j+1}).
\ee
The first sum describes each individual transmon as an anharmonic oscillator with creation/annihilation operators $b_j^\dagger$ and $b_j$.  This is valid in the limit in which we work, where the Josephson energy of each qubit dominates the capacitive charging energy by roughly two orders of magnitude.  Capacitive coupling, or dipole-dipole interaction, between the qubits gives rise to the second term in Eq.~(\ref{H_array}), with $J$ being the hopping amplitude for an excitation to jump from one qubit to the next.  Here, for simplicity, we ignore couplings beyond nearest-neighbors, because the dipole interaction scales inversely with the cube of the qubit-qubit distance, making next-nearest-neighbor coupling amplitudes close to an order of magnitude lower than corresponding nearest-neighbor values.

For a derivation of Eq.~(\ref{H_array}) via circuit quantization, see the supplementary materials of either~\cite{PhysRevLett.110.030601} or ~\cite{DalArxiv}.  These derivations specialize to the lowest two transmon levels, but considering higher levels is straightforward. In this work, we work in a parameter regime where the qubit-qubit coupling is much lower than the qubit frequencies; in such a regime the rotating-wave approximation is well-justified, and we use it here to obtain the Bose-Hubbard Hamiltonian. This Hamiltonian has U(1) invariance and therefore preserves the number of excitations.

The array is coupled to a single-mode  cavity $ H_{\rm cav}=\hbar \omega_c a^\dagger a$ (with $a^\dagger$ and $a$ the creation/destruction operators) via an interaction term
$H_{\rm int}=\hbar \sum_{j=1}^L g_j(b_j  a^\dagger +b_j^\dagger a)$. The coupling $g_j$ between qubit $j$ and the cavity depends on the product of the strength of the electromagnetic field of the cavity mode at the location of qubit $j$ and the transition dipole moment of that qubit, and hence varies from qubit to qubit.  For example, in a setup such as the one used in the main text, where the array lies roughly in the center of the cavity, the qubit-cavity coupling to the lowest cavity mode will be largest in the center of the array.

 Since the operators which create and destroy transmon excitations satisfy the bosonic commutation relation $[b_j, b_j^\dagger]=1$, it is natural to view these excitations as bosons on a lattice, with each transmon embodying a lattice site.  Each transmon has a negative anharmonicity $\alpha_j <  0$, representing an effective on-site interaction between the bosonic excitations.  In the case where the transmon anharmonicity is uniform across the array, close to what we realize in the experiment, the array Hamiltonian in Eq.~(\ref{H_array}) has the familiar form of a Bose-Hubbard model in the regime of attractive interactions (here $\alpha $ plays the role of the Bose-Hubbard parameter $U$).


To complete the description of this circuit-QED system we add the drive term, representing an external classical drive applied to the cavity: $H_{\rm drive}=\hbar \epsilon(t)(a^\dagger e^{-i \omega_d t}+a e^{i \omega_d t})$, where
$\epsilon(t)$ is the strength of the external drive and $\omega_d$ the drive frequency. The full Hamiltonian of the system qubit-array and cavity is the sum of all the aforementioned terms:
\be\label{H_full}
H=H_{\rm array}+H_{\rm cav}+H_{\rm int}+H_{\rm drive}.
\ee

This Hamiltonian describes all aspects of the array-cavity system important to our experiment, including the cooling via engineered dissipation, as we will show shortly.  To see this, we cast the Hamiltonian in a simpler form via a few standard transformations.  First, we move to the rotating frame of the drive by applying the unitary transformation: $U=e^{i\hbar \omega_d t(a^\dagger a+\sum_j b_j^\dagger b_j)}$. The transformed Hamiltonian $\tilde{H}=U H U^\dagger -i \hbar U \partial_t U^\dagger$\footnote{Recall that using the Baker-Campbell-Hausdorff formula we have:
$\tilde{H}=U H U^\dagger=e^{X}H e^{-X}=H+[X,H]+\frac{1}{2!}[X,[X,H]]+\frac{1}{3!}[X,[X,[X,H]]]\dots $ } takes the form:
\be\label{H}
\tilde{H}=\hbar \sum_{j=1}^L \left(\Delta_{j} b_j^\dagger b_j+\frac{\alpha_j}{2}  b_j^\dagger b_j^\dagger b_j b_j \right)+\hbar J\sum_{j=1}^{L-1}(b_{j+1}^\dagger b_j+b_j^\dagger b_{j+1})+
\hbar\Delta_c a^\dagger a+\hbar \sum_{j=1}^L g_j(b_j  a^\dagger +b_j^\dagger a)+\hbar\epsilon(t)(a^\dagger +a ),
\ee
where we have introduced the detunings: $\Delta_c=\omega_c-\omega_d$ and $\Delta_{j}=\omega_j-\omega_d$.  In essence, this rotation gives new effective qubit and cavity frequencies while removing the oscillating time-dependence on the cavity operators in the drive term.

Ignoring for a moment  both the drive and the anharmonic transmon terms in $\tilde{H}$, the array-cavity system is simply a set of coupled harmonic oscillators which exhibits its own set of normal modes.  To find these modes, we write the quadratic terms of the Hamiltonian, $H_0$, in matrix form:
\beq\label{H0}
H_0&=&\hbar \sum_{j=1}^L \Delta_{j} b_j^\dagger b_j+\hbar J\sum_{j=1}^{L-1}(b_{j+1}^\dagger b_j+b_j^\dagger b_{j+1})+
\hbar\Delta_c a^\dagger a+\hbar \sum_{j=1}^L g_j(b_j  a^\dagger +b_j^\dagger a) \\
&=& \hbar\mathbf{v}^\dagger \mathcal{H}_0 \mathbf{v},
\eeq
where we have introduced the $L+1$-component vector $\mathbf{v}=(a, b_1, b_2, \dots b_L)$ consisting of the annihilation operators for each oscillator, and the $(L+1)\times (L+1)$ matrix $\mathcal{H}_0$ representing their frequencies and couplings.  Now the normal modes of the coupled system are found simply by diagonalizing the matrix $\mathcal{H}_0$.  We thus find the matrix $M$ whose columns are the eigenvectors of $\mathcal{H}_0$; then $M^{-1} \mathcal{H}_0 M$ is diagonal.  Calling $N = M^{-1}$, the corresponding change of basis $\mathbf{W} = N \mathbf{v}$ gives the eigenmodes of the system.  Writing these modes explicitly in terms of the matrix elements of $N$ and $M$ will be useful for later purposes---calling the new basis vectors $\mathbf{W}= (A, B_1, B_2,  \dots B_L)$, we have:
\beq
A=N_{00} a+\sum_{l=1}^L N_{0 l} b_{l}, \label{AofN}\\
B_j=N_{j0} a+\sum_{l=1}^L N_{j l} b_{l},\label{BofN}
\eeq
and their inverse transformations:
\beq
a=M_{00} A+\sum_{l=1}^L M_{ 0 l} B_{l},\label{aofN} \\
b_j=M_{j0} A+\sum_{l=1}^L M_{j l} B_{l}.\label{bofN}
\eeq
The corresponding eigenvalues or normal-mode frequencies are denoted by $\lambda_i$. In this new basis $H_0$ is, as expected, a sum of uncoupled harmonic oscillators: $H_0=\hbar \lambda_0 A^\dagger A+\hbar \sum_{j=1}^L \lambda_j B_j^\dagger B_j$.

In the dispersive limit, where $g_j \ll |\omega_{\mathrm{c}} - \omega_{\mathrm{j}}|$, we expect the dressed mode $A$ to largely consist of the cavity mode, with small contributions from each qubit.  Equivalently, this means that $M_{0 l}$ and $N_{0 l}$ are each much smaller than $M_{00}$ and $N_{00}$, which are both close to unity.  Conversely,  the $B_j$ modes are mostly linear combinations of transmon excitations ($b_1, b_2, \dots b_L$) with a small component from the cavity.  In symbols, $|M_{j0}|\ll |M_{jl}|$ and $|N_{j0}|\ll |N_{jl}|$ for $j,l \neq 0$ (see for instance the specific example from the experiment, shown in Eq.~(\ref{N_ex}) and ~(\ref{M_ex})).

In this new basis the drive term becomes: $\epsilon(t)(a^\dagger +a ) = \epsilon(t)\left(M_{00}(A^\dagger +A) +\sum_{l=1}^L M_{0 l} (B^\dagger_{l}+B_{l})\right)$.  Because of the mode mixing, the drive on the original cavity-mode $a$ now excites all of the normal modes $\{A, \{B_\mathrm{j}\}\}$.  However, since $M_{0l} \ll M_{00}$, we can neglect the $B_j$ terms in this operator unless the drive is close to resonance with one of the qubit-like modes.  In that case, the coefficients $M_{0l}$ determine how responsive the qubit-like mode is to a drive pulse, i.e., which states are dark and which are bright.  We will return to this point later.

Continuing onwards in solving the original Hamiltonian, we now put back the anharmonic transmon terms which we neglected when finding the linear eigenmodes.  In terms of the new normal-mode operators, the anharmonic term is
\beq
\sum_{j=1}^L\frac{\alpha_j }{2} b_j^\dagger b_j^\dagger b_j b_j= \frac{1}{2}\sum_{j=1}^L \alpha_j (M_{j0} A^\dagger+\sum_{l=1}^L M_{j l} B^\dagger_{l})(M_{j0} A^\dagger+\sum_{p=1}^L M_{jp} B^\dagger_{p})(M_{j0} A+\sum_{q=1}^L M_{jq} B_{q})(M_{j0} A+\sum_{s=1}^L M_{js} B_{s}).
\eeq
Invoking the rotating wave approximation to neglect terms which do not conserve excitation number, the remaining terms can be grouped into three categories:
\begin{itemize}
\item self-Kerr corrections to the qubit-like operators:
$
\sum_{lpqs=1}^L \mu_{lpqs}B^\dagger_l B^\dagger_p B_q  B_s$,

where we have defined the tensor:
\be
\mu_{lpqs}=\sum_{j=1}^L \alpha_j M_{jl}M_{jp}M_{jq}M_{js},
\ee
\item a self-Kerr correction for the cavity:
$
\Pi_0  A^\dagger A^\dagger A A$,

where we have defined the constant:
\be
\Pi_0=\sum_{j=1}^L  \alpha_j M_{j0}^4,
\ee
\item and a cross-Kerr term that couples together the $A$ and  $B_j$ operators:
$
4 A^\dagger A  \sum_{lp}  \eta_{lp} B^\dagger_l B_p$,

where we have defined the tensor:
\be \label{eta}
\eta_{lp}=\sum_{j=1}^L  \alpha_j M_{j0}^2 M_{jl}M_{jp}.
\ee
\end{itemize}

Now we have the full Hamiltonian in the new basis $\mathbf{W}$:
\beq
\tilde{H}_W& =& \hbar \lambda_0 A^\dagger A+\frac{\hbar}{2}\Pi_0  A^\dagger A^\dagger A A+\hbar \sum_{j=1}^L \lambda_j B_j^\dagger B_j+\frac{\hbar}{2}\sum_{lpqs=1}^3 \mu_{lpqs}B^\dagger_l B^\dagger_p B_q  B_s+2 \hbar A^\dagger A  \sum_{lp}  \eta_{lp} B^\dagger_l B_p\nonumber \\
& & +\hbar \epsilon(t)M_{00}(A^\dagger +A) +\hbar \epsilon(t)\sum_{l=1}^L M_{0 l} (B^\dagger_{l}+B_{l}).
\eeq
Given our assumption  $|M_{j0}|\ll |M_{jl}|$ for $l,j \neq 0$ we have: $\Pi_0\ll \eta_{lp} \ll\mu_{lpqs}$.  In other words, in this new dressed-state basis $\mathbf{W}$ we have:
 [qubit self-Kerr] $\gg$ [cross-Kerr]  $\gg$  [cavity self-Kerr].  For simplicity we henceforth neglect the cavity self-Kerr term $ \Pi_0  A^\dagger A^\dagger A A$.

Next, we displace the new cavity-like operator $A$ and write it as a classical part and a small quantum correction:
\be
A=\bar{A}(t)+D,
\ee
by applying the unitary transformation: $U_D=\exp[\bar{A}(t)A^\dagger-\bar{A}^\star(t)A]$.
Under this transformation $U_D A U_D^\dagger=A-\bar{A(t)}=D$ and
 the Hamiltonian $\tilde{H}_W$ transforms as:
\be
\tilde{H}_{W,D}=U_D \tilde{H}_{W} U_D^\dagger-i \hbar U_D \partial_t U_D^\dagger.
\ee
By direct substitution of $A=\bar{A}(t)+D$ into the Hamiltonian we find:
\be
A^\dagger A =  (\bar{A}^\star +D^\dagger)(\bar{A} +D)=|\bar{A}|^2+(\bar{A}D^\dagger+\bar{A}^\star D)+D^\dagger D
\ee
and therefore:
\beq
\tilde{H}_{W,D}& =&\hbar \lambda_0 \left[|\bar{A}|^2+(\bar{A}D^\dagger+\bar{A}^\star D)+D^\dagger D\right] +\hbar \sum_{j=1}^L\lambda_j B_j^\dagger B_j+\frac{\hbar}{2}\sum_{lpqs=1}^L \mu_{lpqs}B^\dagger_l B^\dagger_p B_q  B_s+\hbar \epsilon(t)M_{00}(D^\dagger +D+\bar{A}^\star+\bar{A})\nonumber\\
& +& 2\hbar \left[|\bar{A}|^2+(\bar{A}D^\dagger+\bar{A}^\star D)+D^\dagger D\right]\sum_{lp}  \eta_{lp} B^\dagger_l B_p-i (\dot{\bar{A(t)}} A^\dagger- \dot{\bar{A(t)^\star}} A)+\hbar \epsilon(t)\sum_{l=1}^L M_{0 l} (B^\dagger_{l}+B_{l}).
\eeq
We choose $\bar A(t)$ by requiring the terms linear in $D$ and $D^\dagger$ to vanish
\be\label{dispA}
\hbar \lambda_0 \bar{A}(t)+(2\hbar \sum_{lp}  \eta_{lp} B^\dagger_l B_p)\bar{A}(t)+\hbar \epsilon(t)M_{00} =-i \hbar\dot{\bar{A}}(t),
\ee
so that  we can eliminate the terms involving the drive. Eliminating the drive is equivalent to moving to a frame where the cavity evolves according to the usual  Heisenberg equation $\dot{A}=\frac{i}{\hbar}[H,A]-\frac{\kappa}{2}M_{00} A$, and therefore solving for the classical part
  $\bar{A}(t)$ is equivalent to  Eq.~(\ref{dispA}).
Since $\eta_{lp} \ll \lambda_0$ we neglect that term and consider only:
\be
\lambda_0 \bar{A}(t)+\epsilon(t)M_{00} =-i \dot{\bar{A}}(t).
\ee
In the stationary case $\dot{\bar{A}}(t)=0$  the solution is:
\be\label{stationary_case}
\bar{A}(t)=\frac{-\epsilon(t) M_{00}}{\lambda_0-i \frac{\kappa}{2}M_{00}},
\ee
where we have included the correction due to the cavity damping rate $\kappa$.

The final Hamiltonian then becomes:
\beq\label{Hcool}
\tilde{H}_{W,D}& =&\hbar \lambda_0 \left[|\bar{A}|^2+D^\dagger D\right] +\hbar \sum_{j=1}^L \lambda_j B_j^\dagger B_j+\frac{\hbar}{2}\sum_{lpqs=1}^L \mu_{lpqs}B^\dagger_l B^\dagger_p B_q  B_s+\hbar \epsilon(t)\sum_{l=1}^L M_{0 l} (B^\dagger_{l}+B_{l})\nonumber\\
& +& 2\hbar \left[|\bar{A}|^2+(\bar{A}D^\dagger+\bar{A}^\star D)+D^\dagger D\right]\sum_{lp}  \eta_{lp} B^\dagger_l B_p.
\eeq
Ignoring the constant offset term $\hbar\lambda_0|\bar{A}|^2$, we break the Hamiltonian up into the following pieces.  We call $H_D$ the term involving only the dressed cavity operator:
\be
H_D=\hbar \lambda_0 D^\dagger D;
\ee
$H_B$ the one involving only the dressed qubit operators:
\be\label{HB}
H_B=\hbar \sum_{j=1}^L \lambda_j B_j^\dagger B_j+\frac{\hbar}{2}\sum_{lpqs=1}^L \mu_{lpqs}B^\dagger_l B^\dagger_p B_q  B_s;
\ee
with  $H_{D,B}$ the interaction between dressed cavity and dressed qubits:
\be\label{H_DB}
H_{D,B}=2\hbar \left[|\bar{A}|^2+(\bar{A}D^\dagger+\bar{A}^\star D)+D^\dagger D\right]\sum_{lp}  \eta_{lp} B^\dagger_l B_p.
\ee
Finally, we have $H_{B,\rm{drive}}$, the driving term on the dressed qubit operators:
\be\label{H_B_drive}
H_{B,\rm{drive}}=\hbar \epsilon(t)\sum_{l=1}^L M_{0 l} (B^\dagger_{l}+B_{l}).
\ee
The final Hamiltonian is the sum of all of these terms:
\be \label{Heff}
H^{\rm eff}=H_D +H_B+H_{D,B}+H_{B,\rm{drive}}.
\ee

\subparagraph{Cooling} The term $H_{D,B}$ is the one we are most interested in.
Depending on the detuning of the drive, this term can act either as a cooling term or as a term that induces a state-dependent shift of the cavity resonance and therefore allows one to detect the state of the system from a homodyne measurement. It is important to notice that this term preserves the U(1) invariance of the Hamlitonian $H_B$ and therefore conserves the total qubit excitation number $N$, thus allowing for cooling within a given excitation-number manifold.

Let us focus on the cooling process first: we assume the loss rate of photons from the cavity to be high with respect to the dynamics involving
  $H_{D,B}$. Therefore,  we neglect the terms $D^\dagger D$ and $\bar{A}^\star D$
and we call  the remaining operator the cooling term:
\be\label{Vcool}
V_{\rm cool}=2 \hbar \bar{A}(t) D^\dagger \sum_{lp}  \eta_{lp} B^\dagger_l B_p=2 \hbar \bar{A}(t) D^\dagger \mathcal{O}_B.
\ee
Within our approximation, the dissipation makes $V_{\rm cool}^\dagger$ ineffective.
In Eq.~(\ref{Vcool})  we have defined the operator:
\be\label{Ob_ope}
\mathcal{O}_B=\sum_{lp}  \eta_{lp} B^\dagger_l B_p.
\ee
If the pump is red-detuned from the cavity, then
the cooling operator $V_{\rm cool}$ scatters one qubit excitation to a lower energy state by creating a photon with higher frequency than the incoming one (strictly speaking, it scatters a pump photon up to the cavity frequency). Therefore the total number of qubit excitations is conserved, but some excess energy has been transferred from the qubit array to a cavity photon.

To estimate the rate of cooling between two states we use the Fermi's Golden Rule:
\be\label{FermiGolden}
\Gamma_{\rm cool}=\frac{2 \pi}{\hbar} \sum_f |\langle \Psi_f | V_{\rm cool} | \Psi_i\rangle|^2 \delta(E_i-E_f).
\ee
In the Hilbert space of the dressed excitations $|\rm qubits \rangle \otimes |\rm cavity \rangle$, the initial state is $| \Psi_i\rangle=|\psi_i, 0\rangle$. We consider the qubits  initially in an excited state $\psi_i$ with energy $E_i$ and  no photons in the cavity. The final state instead  is $| \Psi_f\rangle=|\psi_f, 1\rangle$, where the qubits are in a lower energy state  $\psi_f$ with energy $E_f$,  and the excess energy $E_i-E_f$ is carried away by a new photon. Therefore we have:
\beq
\Gamma_{i\to f}& =& \frac{2 \pi}{\hbar} (2 \hbar \bar{A}(t))^2  \sum_q \langle \psi_i, 0 | \mathcal{O}_B^\dagger D  | \psi_f, 1\rangle \langle \psi_f, 1 | D^\dagger \mathcal{O}_B  | \psi_i, 0\rangle \delta(E_i-E_f-\epsilon_q)\nonumber\\
& = & (2 \pi \hbar) (2 \bar{A}(t))^2 \langle \psi_i| \mathcal{O}_B^\dagger | \psi_f\rangle \langle \psi_f| \mathcal{O}_B | \psi_i\rangle \sum_q \langle 0| D | 1\rangle \langle 1| D^\dagger | 0\rangle \delta(E_i-E_f-\epsilon_q)\nonumber\\
& = &   (2  \bar{A}(t))^2|M_{if}|^2 \int_{- \infty}^{+\infty} dt \sum_q  \langle 0| D | 1\rangle \langle 1| D^\dagger | 0\rangle e^{\frac{i t}{\hbar}(E_i-E_f-\epsilon_q)}\nonumber\\
& = &   (2  \bar{A}(t))^2|M_{if}|^2 \int_{- \infty}^{+\infty} dt \; e^{i t (\omega_i-\omega_f)}
 \sum_q  \langle 0| D(t) | 1\rangle \langle 1| D^\dagger | 0\rangle \nonumber\\
 & = & (2  \bar{A}(t))^2  |M_{if}|^2 \int_{- \infty}^{+\infty} dt \; e^{i t (\omega_i-\omega_f)}
  \langle 0| D(t)  D^\dagger | 0\rangle \nonumber\\
  & = & (2 \bar{A}(t))^2  |M_{if}|^2 S_{DD}(\omega_i-\omega_f),
  \label{RateCooling}
 \eeq
 where $M_{if}=\langle \psi_f| \mathcal{O}_B | \psi_i\rangle$ and
 \be
 S_{DD}(\omega)=\frac{\kappa}{(\omega-\Delta_c)^2+(\kappa/2)^2}
 \ee
 is the  spectral density of the cavity field fluctuations. The square of the classical drive amplitude $\bar{A}(t)^2$ is equal to the average photon number in the cavity: $\bar{A}(t)^2=\bar{n}$, so we see that the cooling rate is linear in the photon number:
 \be\label{RateCooling_short}
 \Gamma_{i\to f}= 4 \bar{n}|M_{if}|^2 \frac{\kappa}{(\omega_i-\omega_f-\Delta_c)^2+(\kappa/2)^2}.
 \ee
 In the stationary case (Eq.~(\ref{stationary_case})) $\bar{A}(t)\sim \epsilon(t)$, and therefore the drive power is directly proportional to the photon number $P_{\rm in}\propto \bar{n}$, giving the result contained in the main text (Eq.(4)) that: $\Gamma_{\rm cool}  \propto P_{\rm in}$.

The transition rates just derived are perturbative, since the derivation relied on Fermi's Golden rule.  Thus they are valid only in the limit that the rate of transitions is small with respect to the decay rate of the cavity:
\be
\Gamma_{\rm cool} < \kappa.
\ee
Beyond this regime the perturbative description of the cooling process is not appropriate anymore because
the effect of the $V_{\rm cool}^\dagger$ term will no longer be negligible, and will induce other processes besides the cooling \cite{Murch_PRL}.

\subparagraph{Stark shift}
For low intracavity photon number, the frequencies $\omega_i$ and $\omega_f$ used above in Eqs.~(\ref{RateCooling}) and~(\ref{RateCooling_short}) are simply the bare eigenvalues  of the Hamiltonian in Eq.~(\ref{HB}) that we diagonalized to find the eigenmodes.  For higher photon number, the first term $2\hbar \,|\bar{A}|^2\sum_{lp}  \eta_{lp} B^\dagger_l B_p$ of the Hamiltonian $H_{D,B}$ (Eq.~(\ref{H_DB})) starts to be significant; this term contributes  a
Stark shift to the energies of the eigenstates. We consider this Stark shift perturbatively and define a photon-number dependent frequency as:
\be
\omega_i(\bar{n})=\omega_i^0+2 \hbar\,\bar{n}\langle \psi_i | \mathcal{O}_B|\psi_i\rangle.
\ee
For the low photon number used in the experiment (at most five photons at steady state in the cavity), this perturbative correction is a good approximation to the exact solution.

\subparagraph{Measurement}
 By performing a homodyne measurement, we can measure the  shift of the cavity frequency depending on the state of the array:
 \be\label{cavity_pull}
  \left[ \lambda_0 +2  \sum_{lp}  \eta_{lp} B^\dagger_l B_p\right] D^\dagger D.
  \ee
 Therefore we introduce the operator $\chi$ describing the cavity pull:
 \be\label{chi_shift_op}
 \chi= 2\sum_{lp}  \eta_{lp} B^\dagger_l B_p=2 \mathcal{O}_B,
 \ee
The expectation value of this operator on a generic state of the array $S$ constitutes the observable that identifies that state:
\be\label{chi_shift}
 \chi_S= 2\langle \mathcal{O}_B\rangle_S =2\langle \sum_{lp}  \eta_{lp} B^\dagger_l B_p \rangle_S.
  \ee
 We point out an important difference in the behavior of the operator $\mathcal{O}_B$ during the cooling process versus during the measurement.
 During the cooling process the pump drive is centered at a red detuning  $\Delta_c=\omega_i-\omega_f$ given by the energy difference
 between the two states, initial and final, constituting the cooling transition.
 The operator $\mathcal{O}_B$ then will induce transitions between states whose energy difference lies within a bandwidth $\sim\kappa$ centered on $\Delta_c$.
During the readout, instead, the drive  is at the cavity frequency, i.e.\ at zero detuning, so to a good approximation we can neglect terms
rotating faster than the linewidth $\kappa$ of the cavity. For small lattice sizes and big coupling $J$, as it is the case in our experiment with $L=3$, the excitation energies are big compared to $\kappa$ and therefore the operator $\chi$ will not induce transitions between different states during the readout, and the measurement can then be considered quantum nondemolition (QND).

\subsection{Long array limit}
For long arrays, $L\gg 1$, and small hopping strength $J$, the energy levels will become closely spaced and therefore eventually the measurement will not be QND anymore.  Once there exist pairs of states with energy difference $\Delta E$ less than $\sim\kappa$, the measurement drive will (generically) cause transitions between them.  This same physics will adversely affect the cooling process that we have discussed.
The cooling process  implements the removal of an amount of energy $\Delta E=\hbar(\omega_i -\omega_f)$ ($\omega_i> \omega_f$) that is well-defined up to the cavity linewidth $\kappa$. $\Delta E\gg \kappa$ is easily achievable in small arrays. In the limit of long arrays, eventually $\Delta E \sim \kappa$ or smaller and both Stokes and anti-Stokes processes become possible and the (non-equilibrium) quantum bath no longer has zero effective temperature \cite{SteveRMP}.

Consider the simplest case in which all of the quits have uniform frequency and are coupled only to nearest-neighbors.  Neglecting initially the anharmonic terms, i.e.\ imagining a chain of coupled linear harmonic oscillators gives the simple Hamiltonian
\be\label{H_tightb}
H_{\rm array}=\hbar \omega_0 \sum_{j=1}^L  b_j^\dagger b_j+\hbar J\sum_{j=1}^{L-1}(b_{j+1}^\dagger b_j+b_j^\dagger b_{j+1}).
\ee
This chain of oscillators exhibits normal modes with excitation amplitudes varying sinusoidally along the array.  Or, in the language of excitations as bosonic particles, this is a tight-binding model of free bosons on a lattice, which is diagonalized by introducing modes of the form:
\be
B_n=\sqrt{\frac{2}{L+1}} \sum_{j=1}^L\sin(k_n j) b_j, \qquad k_n=\frac{\pi}{L+1} n, \quad n=1,2,\dots, L.
\ee
In this form the Hamiltonian becomes
\be\label{Hcos}
H_{\rm array}=\hbar \sum_{n=1}^L [\omega_0+2J\cos(k_n)] B^\dagger_n B_n.
\ee
Note that unlike in the previous derivation we did not include the coupling to the cavity when diagonalizing the quadratic Hamiltonian.  Including these terms we now obtain
\begin{eqnarray}\label{HCavQub}
H &=& \hbar \sum_{n=1}^L [\omega_0+2J\cos(k_n)] B^\dagger_n B_n + \hbar\omega_c a^\dagger a + \hbar \sum_{j=1}^L g_j\left(b_j a^\dagger + b_j^\dagger a\right)\\
&=&\hbar \sum_{n=1}^L [\omega_0+2J\cos(k_n)] B^\dagger_n B_n + \hbar\omega_c a^\dagger a + \hbar \sum_{m=1}^L \xi_m\left(B_m a^\dagger + B_m^\dagger a\right)
\end{eqnarray}	
where
\be
\xi_m= \sqrt{\frac{2}{L+1}}\sum_{j=1}^L g_j\sin(k_mj).  
\ee
The quartic interaction term becomes in this basis
\begin{eqnarray}
H_{\mathrm int} &=& \frac{1}{2}\sum_{j=1}^L \alpha_j b^\dagger_j b^\dagger_j b_j b_j\\
&=& \sum_{m,n,p,q=1}\Xi_{mnpq}B^\dagger_mB^\dagger_nB_pB_q,
\end{eqnarray}
where
\be
\Xi_{mnpq}=\frac{2}{(L+1)^2} \sum_{j =1}^L \alpha_j\sin(k_mj)\sin(k_nj)\sin(k_pj)\sin(k_qj).  
\ee

Proceeding as before we can diagonalize the quadratic part of the Hamiltonian in the presence of a drive on the cavity and extract the cooling operator defined in Eq.~(\ref{Vcool}) and Eq.~(\ref{Ob_ope}) and hence determine the cooling matrix elements $M_{if}$.   We will not present those details here but rather simply note certain qualitative features of the result that can be seen more easily in the present version of the derivation where we have delayed inclusion of the cavity coupling until after diagonalization of the array Hamiltonian.
If $g_j$ varies slowly and smoothly with position then $\xi_m$ will be large only for small $m$.  It follows that the Raman scattering process does not permit large changes of (quasi-) momentum and the matrix $\eta_{lp}$ in Eq.~(\ref{Ob_ope}) will be non-zero only near the diagonal.  If only small momentum changes are permitted then only small energy changes are permitted and the cooling will be weak.   If on the other hand, the set of $\{g_j\}$ has low symmetry then the $\{\xi_m\}$, and hence the transition matrix elements $M_{if}$ derived from them, are not constrained by symmetry and are generically non-zero.

\begin{figure}[htbp]
\begin{center}
\includegraphics[width=.30\textwidth]{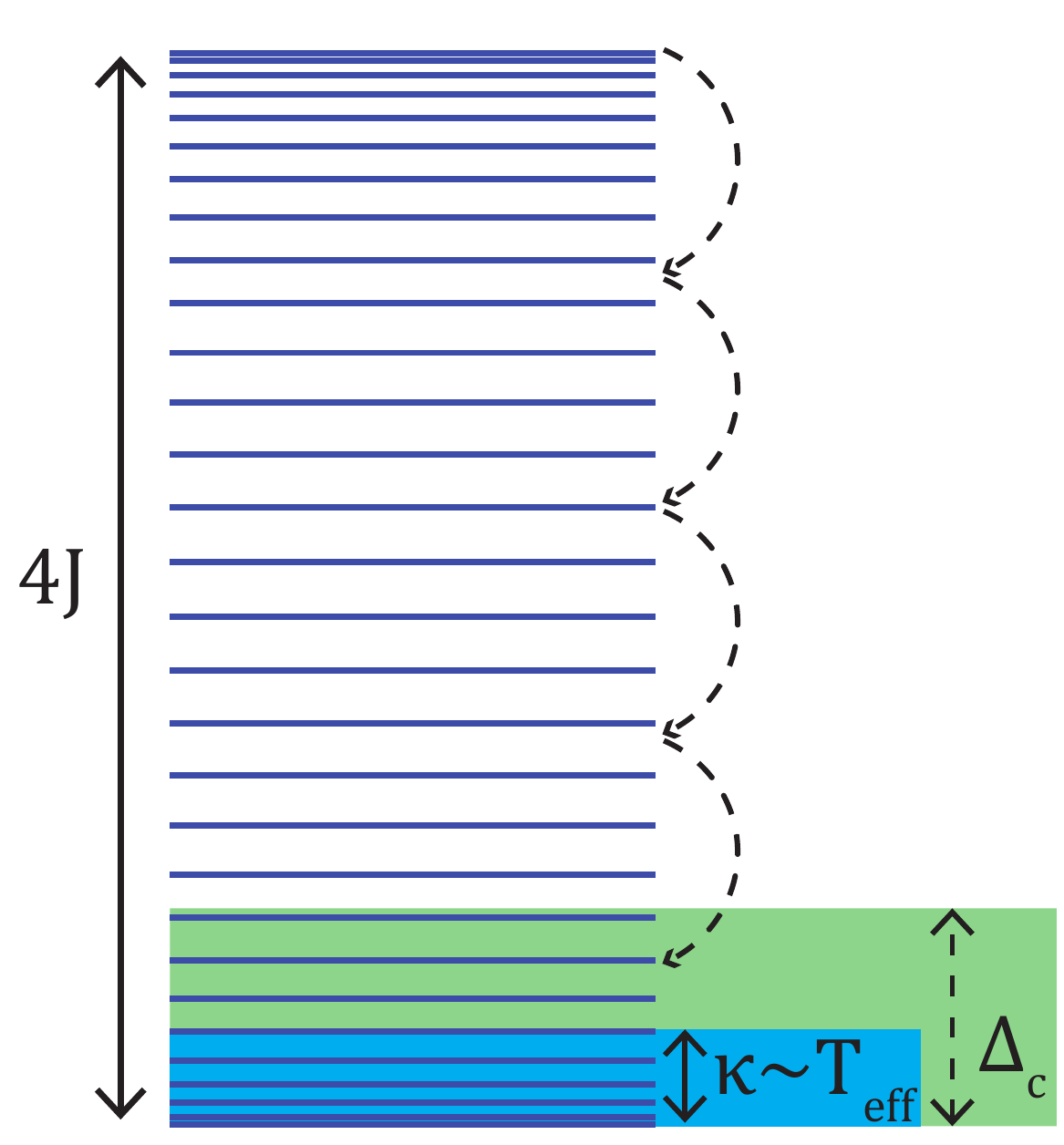}
\caption{Illustration of the temperature limit for the single excitation subspace of a degenerate L-site chain. To be able to cool to the ground state we require  $J(\frac{\pi}{L+1})^2>\kappa$.}   
\label{Teff}
\end{center}
\end{figure}

We now discuss a protocol to cool towards the lower energy eigenstates of  a general L-site array. We illustrate the basic idea using the single-excitation manifold as an example.
Based on the band structure of the 1D tight-binding model, we know that the single-excitation manifold energies are spread around the bare qubit frequency $\omega_0$  with a total width of $ 4 J$.
The levels will be more densely spaced at the top and bottom of the band (see Fig.~\ref{Teff}) due to the van Hove singularities in the extremes of the tight-binding band structure.

For simplicity we assume that all cooling matrix elements are generically non-zero.   Suppose we start from a high-energy state $E_i$  and we set the drive such that the cavity detuning $\kappa<\Delta_c< 4J$ can cool between transitions within the bandwidth. As illustrated in Fig.~\ref{Teff}, a cascade of several cooling steps brings the initial energy down to a final value equal to $E_0^{(f)}=E_0\mod(\Delta_c)<\Delta_c$, at which point no more cooling steps can occur. To lower the energy further, we can slowly decrease the detuning $\Delta_c$ towards $\sim\kappa$. We will be able to cool to the lowest energy state if  $J(\frac{\pi}{L+1})^2>\kappa$. Otherwise this process will cool down the N-particle subspace to an effective temperature of $T_{\mathrm{eff}} \sim \kappa$ \cite{SteveRMP}.

It may occur that the above procedure fails because some particular matrix element in the cascade vanishes.
This can be overcome by scanning cooling pump detuning  $\Delta_c$ down and up multiple times to find a transition that allows escape from the trapped state. 

The cooling rate calculations used above for illustrative purposes are straightforward to carry out within the manifold of single excitations.   Higher excitation manifolds will exhibit more complex and interesting dynamics because as energy is removed from the system by Raman processes, boson-boson collisions will further relax the particle distribution function.  The dynamics will depend importantly on whether or not the underlying model is integrable.  These considerations are well beyond the scope of the present experimental state-of-the-art, but could well become important in future realizations of quantum simulators.

\subsection{Three-qubit array used in the experiment}
We now specialize the above results to the case of $L=3$ sites.
The Hamiltonian  $H_0$ in  Eq.~(\ref{H0}) is:
\be
H_0=\hbar \sum_{j=1}^3 \Delta_{j} b_j^\dagger b_j+\hbar J\sum_{j=1}^{2}(b_{j+1}^\dagger b_j+b_j^\dagger b_{j+1})+\hbar J_{\rm 13} (b_{1}^\dagger b_3+b_3^\dagger b_1)+
\hbar\Delta_c a^\dagger a+\hbar \sum_{j=1}^3 g_j(b_j  a^\dagger +b_j^\dagger a).
\ee
where we have also  included the  next-nearest-neighbor coupling $J_{\rm 13}$ between the 1st and 3rd sites.
This Hamiltonian has the matrix form:
\be
\mathcal{H}_0=\hbar \left(\begin{array}{cccc}\Delta_c & g_1 & g_2 &g_3\\  g_1 & \Delta_1 & J & J_{13}\\ g_2 & J & \Delta_2 & J\\ g_3 & J_{13} & J & \Delta_3   \end{array} \right)
\ee
acting on a four-component vector $\mathbf{v}=(a, b_1, b_2, b_3)$, so that $H_0=\mathbf{v}^\dagger \mathcal{H}_0 \mathbf{v}$.
The matrix $N$ defined in Eqs.~(\ref{AofN}) and (\ref{BofN})  for the experimental parameters given
in Section \ref{Device_param} is:
\be\label{N_ex}
N=\left(\begin{array}{cccc}
-0.986 &  -0.073 & -0.126 & -0.077 \\
 -0.163 &  0.426 & 0.690 & 0.561\\
0.014 & -0.680 & -0.158 & 0.716\\
 -0.013 & -0.592 & 0.695 & -0.408
 \end{array} \right).
\ee
The rows of this matrix are the coefficients of the new dressed operators   $(A, B_1, B_2, B_3)$  expressed in the old basis $(a, b_1, b_2, b_3)$. As expected $A\approx a$ remains an operator that is mostly cavity-like, similarly the $B_j$ operators are mostly linear combination of qubits operators.
This is ensured by the condition $g_j/(\omega_j-\omega_c) \ll 1$ for each $j$. In the experimental setup we have:  $g_{1,3}/(\omega_{1,3}-\omega_c)\sim 0.06$ and  $g_{2}/(\omega_2-\omega_c) \sim 0.12$.

If the couplings $g_1$ and $g_3$ were identical, the model would be symmetric with respect to the exchange of $b_1 \leftrightarrow b_3$, and the operators $A, B_1, B_3$ would  be even under such transformation, while $B_2$ would be odd,
transforming as $B_2 \leftrightarrow -B_2$.
In the experimental case, $g_1$ and $g_3$ are close but not identical and similarly, the qubit frequencies $\omega_1$ and $\omega_3$ are slightly different. This symmetry is thus broken in the experiment, but only weakly. We see this, for instance, in the matrix $N$ that is almost, but not completely, invariant under exchange of $b_1 \leftrightarrow b_3$.

The matrix $M=N^{-1}$, that according to Eqs.~(\ref{aofN}) and (\ref{bofN}) has as elements the decomposition of the original  operators $(a, b_1, b_2, b_3)$ in the new basis $(A, B_1, B_2, B_3)$, is:
\be \label{M_ex}
M=\left(\begin{array}{cccc}
-0.986 & -0.164 & 0.014 & -0.013\\
 -0.073 & 0.426 & -0.680 & -0.592\\
 -0.126 & 0.690 & -0.158 & 0.694\\
 -0.077 & 0.561 & 0.716 & -0.410\\
 \end{array} \right).
\ee
Inspecting the elements of this matrix confirms that for this range of parameters our assumptions  $M_{00}\approx 1$ and $M_{0l}\ll 1$ are valid.

The tensor $\eta$ in Eq.~(\ref{eta}) is:
\be
\eta=2 \pi  \times  \left(\begin{array}{ccc}
-2.422 & 0.227 & -1.241\\
 0.227 & -1.279 & 0.338 \\
 -1.241 & 0.338 & -2.445\\
 \end{array} \right) \rm MHz.
\ee

The fact that the off-diagonal elements in the above matrix involving the operator $B_2$ ($\eta_{12}$ and $\eta_{23}$) are  relatively small is a manifestation of the near-symmetry our system exhibits, described above.  The operator $V_{\rm cool}$ in Eq.~(\ref{Vcool}) cannot connect states with different symmetry (in the perfectly symmetric case $B_1$, $B_3$ would be even, while $B_2$ odd and the elements $\eta_{12}=\eta_{21}$ and $\eta_{23}=\eta_{32}$ would be equal to zero).

\subparagraph{Single-excitation subspace} Let us consider first the eigenstates with only a single excitation in the array.  For these states the nonlinear term in Hamiltonian Eq.~(\ref{HB}) vanishes and we are left with a simple diagonal Hamiltonian:
\be\label{HB_3_one}
H_B=\hbar \sum_{j=1}^L \lambda_j B_j^\dagger B_j.
\ee
A basis $\mathcal{B}$ for the Hilbert space of the one-excitation  manifold
 is the following:
\beq
{\bf |1\rangle}&  = & | 1, 0, 0\rangle=  {B_1^\dagger} |G\rangle \label{basis11},\\
{\bf |2\rangle} &= &| 0,1, 0\rangle=  {B_2^\dagger} |G\rangle\label{basis12},\\
{\bf |3\rangle} &= &| 0, 0, 1\rangle= {B_3^\dagger} |G\rangle\label{basis13}.\\
\eeq
We have introduced here the notation $|m,n,l\rangle \propto (B_1^\dagger)^m  (B_2^\dagger)^n (B_l^\dagger)^l |G\rangle$ as a shorthand way of representing a normalized state with $m$ excitations in the lowest eigenmode, $n$ in the second mode and $l$ in the third mode. Recall that the three modes do not correspond to the three lattice sites;  each operator $B_j$ has in general a nonzero component on each site, as we found in Eq.~(\ref{BofN}).
In this one-manifold basis, the Hamiltonian in Eq.~(\ref{HB_3_one}) is diagonal, and therefore the one-excitation modes $\{|E_i\rangle \text{,~} i\in [1,3]\}$ coincide with this basis: $|E_1\rangle={\bf |1\rangle}$, $|E_2\rangle={\bf |2\rangle}$, $|E_3\rangle={\bf |3\rangle}$, provided that we have ordered them from lowest to highest energy,  $\lambda_1<\lambda_2<\lambda_3$.
Expressed in this basis, the operator $\mathcal{O}_B$ in Eq.~(\ref{Ob_ope}) has the matrix form:
\be
\mathcal{O}_B=\left(\begin{array}{ccc}
\eta_{11} & \eta_{21} & \eta_{31} \\
\eta_{12} & \eta_{22} & \eta_{32} \\
\eta_{13} &  \eta_{12} & \eta_{33}
   \end{array}\right).
\ee
The cooling rate in this manifold, for example, from $\ket{E_3}$ to $\ket{E_1}$ is easily evaluated from  Eq.~(\ref{RateCooling}) as:
\be\label{cool_rate_E}
\Gamma_{E_3\to E_1}=(2 \bar{A}(t))^2  |\eta_{13}|^2 S_{DD}(\lambda_3-\lambda_1).
\ee
The cavity pull, or $\chi$-shift, corresponding to each state is calculated via:
\be\label{chi_E}
\frac{\chi_{ E_i}}{\kappa}=\frac{2  \langle \mathcal{O}_B\rangle_{E_i}}{\kappa}=\frac{2  \eta_{ii}}{\kappa}
\ee

\subparagraph{Two-excitation subspace} Let us consider the manifold consisting of states which have two excitations in the array. We define a basis for this manifold similar to that used for the single-excitation subspace, noting that because of the nonlinear terms, this basis is not an eigenbasis for the Hamiltonian:
\beq
{\bf |1\rangle}&  = & | 2, 0, 0\rangle= \frac{1}{\sqrt{2}} {B_1^\dagger}^2 |G\rangle \label{basis1},\\
{\bf |2\rangle} &= &| 0,2, 0\rangle= \frac{1}{\sqrt{2}} {B_2^\dagger}^2 |G\rangle\label{basis2},\\
{\bf |3\rangle} &= &| 0, 0, 2\rangle= \frac{1}{\sqrt{2}} {B_3^\dagger}^2 |G\rangle\label{basis3},\\
{\bf |4\rangle} &= & | 1, 1, 0\rangle= B_1^\dagger  B_2^\dagger |G\rangle\label{basis4},\\
{\bf |5\rangle} &= & | 0, 1, 1\rangle= B_2^\dagger  B_3^\dagger |G\rangle\label{basis5},\\
{\bf |6\rangle} &= & | 1, 0, 1\rangle= B_1^\dagger  B_3^\dagger |G\rangle\label{basis6}.
\eeq
The Hamiltonian in Eq.~(\ref{HB})
\be
H_B=\hbar \sum_{j=1}^3 \lambda_j B_j^\dagger B_j+\frac{\hbar}{2}\sum_{lpqs=1}^3 \mu_{lpqs}B^\dagger_l B^\dagger_p B_q  B_s
\ee
expressed in the basis $\mathcal{B}$ has the matrix form:
\be\label{matrixM}
\mathcal{H}_B=\left(\begin{array}{cccccc} 2 \lambda_1+ \mu_{1111} &  \mu_{2211} &  \mu_{3311}& \sqrt{2}  \mu_{1211}& \sqrt{2}  \mu_{2311} & \sqrt{2}  \mu_{1311} \\   \mu_{2211}& 2 \lambda_2+ \mu_{2222}  &  \mu_{3322}& \sqrt{2}  \mu_{1222}& \sqrt{2}  \mu_{2322} & \sqrt{2}  \mu_{1322} \\  \mu_{3311}  & \mu_{3322}& 2 \lambda_3+ \mu_{3333}&\sqrt{2}  \mu_{1233}& \sqrt{2}  \mu_{2333} & \sqrt{2}  \mu_{1333} \\
\sqrt{2}  \mu_{1112} & \sqrt{2} \mu_{2212} &\sqrt{2}  \mu_{3312} & \lambda_1+\lambda_2+2  \mu_{1212} & 2  \mu_{2312}  & 2 \mu_{1312}  \\
\sqrt{2}  \mu_{1123} & \sqrt{2}  \mu_{2223} &\sqrt{2}  \mu_{3323} & 2  \mu_{1223}  &\lambda_2+\lambda_3+2  \mu_{2323} &  2  \mu_{1323}  \\
\sqrt{2}  \mu_{1113} & \sqrt{2}  \mu_{2213} &\sqrt{2}  \mu_{3313} & 2  \mu_{1213} &  2  \mu_{2313} &\lambda_1+\lambda_3+2  \mu_{1313}   \\
   \end{array} \right)
\ee
and the operator $\mathcal{O}_B$ in Eq.~(\ref{Ob_ope}):
\be
\mathcal{O}_B=\left(\begin{array}{cccccc} 2 \eta_{11} & 0 & 0& \sqrt{2} \eta_{12} & 0 & \sqrt{2} \eta_{13}\\
0 & 2 \eta_{22} & 0 & \sqrt{2} \eta_{12} & \sqrt{2} \eta_{23} & 0 \\
0 & 0 & 2 \eta_{33} & 0 & \sqrt{2} \eta_{23} & \sqrt{2} \eta_{13} \\
\sqrt{2} \eta_{12} & \sqrt{2} \eta_{12} & 0 & \eta_{11}+\eta_{22} & \eta_{13} & \eta_{23} \\
0 & \sqrt{2} \eta_{23} & \sqrt{2}\eta_{23} & \eta_{13} & \eta_{22}+\eta_{33} & \eta_{12} \\
\sqrt{2} \eta_{13} & 0 & \sqrt{2}\eta_{13} & \eta_{23} & \eta_{12} & \eta_{11}+\eta_{33}
   \end{array} \right).
\ee
To find the eigenstates of the two-excitation manifold we need to diagonalize the matrix:
\be
\mathcal{\tilde{H}}_{B}=\mathcal{H}_B+ 2 \hbar |\bar{A}|^2 \mathcal{O}_B,
\ee
that represents $H_B$ plus the photon-dependent Stark shift $2 \hbar |\bar{A}|^2 \sum_{lp}  \eta_{lp} B^\dagger_l B_p$.
The six eigenstates of the matrix $\mathcal{\tilde{H}}_{B}$ correspond to the six $F$-states introduced in the main text, with corresponding
eigenfrequencies $\epsilon_j$, $j=1, 2, \dots, 6$.

The cooling rate from $\ket{F_i}$ to $\ket{F_j}$ is evaluated from  Eq.~(\ref{RateCooling}) as:
\be\label{cool_rate_F}
\Gamma_{F_i\to F_j}=(2 \bar{A}(t))^2  \langle F_i| \mathcal{O}_B | F_j\rangle|^2 S_{DD}(\epsilon_i-\epsilon_j).
\ee
The cavity pull in a generic $F_j$ state is:
\beq\label{chi_F}
\frac{\chi_{ F_i}}{\kappa}=\frac{2  \langle F_j |\mathcal{O}_B|{F_j}\rangle}{\kappa}
\eeq

\newpage

\section{Summary of theoretical results}

\subsection{Cooling rates}
As shown in Eqs.~(\ref{RateCooling}) and (\ref{cool_rate_E}), (\ref{cool_rate_F}) we can theoretically predict the cooling rate for a specific transition as a function of the average photon number $\bar{n}$ and the detuning of the cooling drive from the cavity resonance:
\be
\Gamma_{i\to j}= 4 \bar{n} |M_{if}|^2 \frac{\kappa}{(\omega_i-\omega_f-\Delta_c)^2+(\kappa/2)^2}.
\ee
As noted above, the frequencies $\omega_i$ and $\omega_f$ in the above equation depend on the photon number via the Stark shift, an effect that was observed experimentally (see Fig. 3b of the main text).  At resonance, the cooling rate per photon number is then simply proportional to the square of the matrix element of the cooling operator on the two states $M_{if}=\langle \psi_f| \mathcal{O}_B | \psi_i\rangle$:
\be\label{cool_per_photon}
\frac{\Gamma^{\rm res}_{i\to j}}{\bar{n}}= \frac{16}{\kappa} |M_{if}|^2.
\ee
In Table \ref{table_cool_rates} we show the cooling rates per photon number for different transitions in the two-excitation subspace and single-excitation subspace, comparing the theoretical predictions with the rates measured in the experiment. The theoretical rates are always within a factor of two of the measured ones, and consistently always higher than the experimental ones. This suggests that the single-mode cavity approximation, or others, in the theoretical model are underestimating other decay processes
which weaken the effective cooling rate accessible experimentally.
\begin{center}
\begin{table}
\caption{Cooling rates $\Gamma_{i\rightarrow f}$ in MHz}
\begin{tabular}{c | c | c | c }
\hline
	$\ket{\psi_i}$ & $\ket{\psi_f}$ & Experiment & Theory \\ \hline
	$\ket{F6}$ & $\ket{F5}$ & 0 & 0 \\
	 & $\ket{F4}$ & 11.6 & 16.1 \\
	 & $\ket{F3}$ & 4.2 & 9.8 \\
	 & $\ket{F2}$ & 0 & 0.28 \\
	 & $\ket{F1}$ & 0 & 0 \\ \hline
	$\ket{F5}$ & $\ket{F4}$ & 0.4 & 0.86 \\
	 & $\ket{F3}$ & 0.5 & 0.75 \\
	 & $\ket{F2}$ & 5.8 & 12.5 \\
	 & $\ket{F1}$ & 0 & 0.2 \\ \hline
	$\ket{F4}$ & $\ket{F3}$ & NA & 10.6 \\
	 & $\ket{F2}$ & 3.1 & 4.2 \\
	 & $\ket{F1}$ & 8.4 & 10.2 \\ \hline
	$\ket{F3}$ & $\ket{F2}$ & 0.6 & 0.66 \\
	 & $\ket{F1}$ & 11 & 20.6 \\ \hline
	$\ket{F2}$ & $\ket{F1}$ & 2.6 & 10.3 \\ \hline
	$\ket{E3}$ & $\ket{E2}$ & 0.54 & 0.52 \\ \hline
	$\ket{E3}$ & $\ket{E1}$ & 13 & 15.5 \\ \hline
	$\ket{E2}$ & $\ket{E1}$ & 0.54 & 1.15 \\ \hline
\end{tabular}
\label{table_cool_rates}
\end{table}
\end{center}

We  now refer back to the simple tight-binding model of free excitations hopping on the lattice (that we have analyzed above in Eq.~(\ref{H_tightb}) and following), we expect this to apply well to the single-excitation manifold where the nonlinear terms are zero. On a three-site lattice the only allowed Fourier modes are: $k_1=\pi/4$, $k_2=\pi/2$ and $k_3=3\pi/4$; these correspond  respectively to the momentum of the states: $E_3$, $E_2$, $E_1$. Given that $J>0$, the state with higher momentum has the lowest energy (here $E_1$ has momentum $k_3=3\pi/4$).  In the case where the couplings obey $g_1=g_3$, they have parity symmetry.  Hence
 cooling can be effective between states $E_3$ and $E_1$ because they have the same spatial parity.  Conversely cooling is highly suppressed in the case of $E_3$ to $E_2$ or $E_2$ to $E_1$ because those transitions require a change of parity.  This is clearly confirmed by the cooling rate experimentally measured between the states of the E-manifold (see lower part of Table \ref{table_cool_rates}).  The cooling rate $\Gamma_{E_3\rightarrow E_1}$ is more than twenty times bigger than $\Gamma_{E_3\rightarrow E_2}$ and $\Gamma_{E_2\rightarrow E_1}$.

\subsection{Calculation of $T_1$ Purcell-limited relaxation time}
We calculate the Purcell relaxation rate as the decay rate of a qubit-array state due to the action of the bare cavity operator $a$; this operator destroys a photon and induces transitions to lower states. In order to estimate the rate for such a process to occur, we compute the overlap between the $\ket{E_i}$ and $\ket{F_j}$ states and the bare cavity mode $a$.
From Eq.~(\ref{BofN}), we know the decomposition of each dressed qubit operator $B_j$ in terms of the bare operators:
\be\label{BofN_bis}
B_j=N_{j0} a+\sum_{l=1}^L N_{j l} b_{l}.
\ee
Since the single-particle states $\ket{E_j}$ are simply given by the dressed operator $B_j$ acting on the vacuum (with no excitations in either qubit or cavity), the overlap between the $\ket{E_j}$ state and the bare cavity mode is simply given by $N_{j0}$.  Then, the Purcell-limited decay rate for this process is the square of this overlap times the cavity decay rate $\kappa$:
\be
R_{E_j \to G}=|\langle G|a |E_j\rangle|^2 \kappa=N_{j0}^2 \kappa, \qquad T_1=\frac{1}{R_{E_j \to G}}.
\ee
The results for the $|E_j\rangle \to |G\rangle$ decay times are shown in the top three rows of Table~\ref{table_T1_th}.

The analysis for the decay rates of the $|F_j\rangle $ states is similar with the decay rate of, say, an initial two-excitation substate to a single-excitation state given by
\be
R_{F_j\to E_i} = |\langle E_i|a |F_j\rangle|^2 \kappa
\ee
To calculate these values we expand $\ket{F_j}$ in the basis $\mathcal{B}$ of states $\{ {\bf | n\rangle}\}_{n\in [1,6]}$ in Eqs.~(\ref{basis1})-(\ref{basis6}):
$|F_j\rangle=\sum_{n=1}^6 c^{F_j}_n {\bf |n\rangle}$.  Then the decay rate to a specific one-excitation state $\ket{E_i}$ is
\be
R_{F_j\to E_i}= \left|\langle  E_i|a \sum_{n=1}^6 c^{F_j}_n {\bf |n\rangle}  \right|^2 \kappa=\left|\sum_{n=1}^6 c^{F_j}_n \langle E_i |a {\bf |n\rangle}  \right|^2 \kappa;
\ee
and the decay rate for a direct decay to the ground state given by
\be
R_{F_j\to G}= \left|{\langle  G} |a^2 \sum_{n=1}^6 c_n {\bf |n\rangle}  \right|^2 \kappa=\left|\sum_{n=1}^6 c_n {\langle  G} |a^2 {\bf |n\rangle}  \right|^2 \kappa.
\ee
The total Purcell-limited decay rate is the sum of the rates given above, with a Purcell-limited $T_1$ time given by the inverse of that sum.  The $T_1$ times found from the above analysis are shown in  Table~\ref{table_T1_th}.
For comparison,  the $T_1$ decay times measured in the experiment are shown in Table \ref{table_T1_exp_down} for the downward rates.  Due to spurious excitation from the sample being at finite temperature, the experiment observed upward transition rates as well.  These spurious excitation rates are shown in Table \ref{table_T1_exp_up}.

In most of the cases, the theoretical prediction for $T_1$ is of the same order of magnitude as the experimentally measured value.  Nevertheless, there are some qualitative discrepancies.  For instance, the theoretical $T_1$ times for the $F$-states are in general shorter than the measured ones. We attribute this to the limitation of the single-mode cavity model which can be shown to predict a shorter life-time than a model which takes into account higher modes of the 3D cavity.  A more mundane discrepancy arises from the fact that as the $\ket{E_1}$ and $\ket{E_2}$ are almost dark states, their dominant decay channel is not Purcell decay via coupling to the cavity, but rather on-chip material losses.

 \begin{center}
 \begin{table}
\caption{ $T_1$ theoretical Purcell-limited decay time in $\mu$s}
\begin{tabular}{c |c| c |c|c|c|c }
\hline
state & $\omega/(2 \pi)$ GHz  & $T_1$(tot) & $T_1(E_1)$ & $T_1(E_2)$& $T_1(E_3)$ & $T_1(G)$  \\
\hline
$E_1$ &  4.61164 & 97 &  & & &  97\\
$E_2$& 4.85539  &  80 & & & & 80\\
$E_3$ & 5.0196  &  0.6 & & & & 0.6 \\
\hline
$F_1$ &  9.11862 &  20 & 32& 201 & 93 & 438\\
$F_2$ &  9.3201&  7.5  & 8.8 & 57 & 705 & $>$1ms \\
$F_3$& 9.48676  & 1.3 & 1.3 & 50 & 212 & $>$1ms \\
$F_4$ & 9.64465  & 1.2  & 1.3 & 69 & 30 & 182\\
$F_5$&  9.7987 &  0.6 & 49 & 0.6 & 159 & $>$1ms \\
$F_6$ &  9.97278 & 0.9 & 78 & $>$1ms & 1.17 & 5.7\\
\hline
\end{tabular}
\label{table_T1_th}
\end{table}
\end{center}

 \begin{center}
 \begin{table}
 \caption{ $T_1$ experimental fitted \emph{downward} decay time  in $\mu$s}
\begin{tabular}{c |c| c |c|c|c|c }
\hline
state & $\omega/(2 \pi)$ GHz  & $T_1$(tot) & $T_1(E_1)$ & $T_1(E_2)$& $T_1(E_3)$ & $T_1(G)$  \\
\hline
$E_1$ &  4.6078 & 28.5 &  & & &  28.5\\
$E_2$ & 4.7854  &  30.5 & & & & 30.5\\
$E_3$ & 5.06916  &  3.2 & & & & 3.2\\
\hline
$F_1$ &  9.1184 &  18.0 & 20.4 & 151 &  & \\
$F_2$ &  9.2592&  15.1  & 30.3 & 30.1 &  &  \\
$F_3$ & 9.4230  & 8.8 & 13.5 & 25.4 &  &  \\
$F_4$ & 9.5618  & 4.6  & 5.3 & 34.6 &  & \\
$F_5$ &  9.7788 &  3.1 & 100.7 & 3.3 & 70.0 & \\
$F_6$ &  10.0539 & 1.5 & 20.6 & 42.6 & 1.6 & \\
\hline
\end{tabular}
\label{table_T1_exp_down}
\end{table}
\end{center}

 \begin{center}
 \begin{table}
 \caption{ $T_1$ experimental fitted \emph{upward} decay time  in $\mu$s}
\begin{tabular}{c |c| c |c|c|c|c }
\hline
state & $\omega/(2 \pi)$ GHz  & $T_1$(tot) & $T_1(E_1)$ & $T_1(E_2)$& $T_1(E_3)$ & $T_1(G)$  \\
\hline
$E_1$ &  4.6078 & 82.4 &  & & &  82.4\\
$E_2$ & 4.7854  &  182.3 & & & & 182.3\\
$E_3$ & 5.06916  &  167.0 & & & & 167.0\\
\hline
$F_1$ &  9.1184 &  104.2 & 104.2 &  &  & \\
$F_2$ &  9.2592&  82.0  & 237.5 & 125.2 &  &  \\
$F_3$ & 9.4230  & 45.3 & 98.1 & 84.2 &  &  \\
$F_4$ & 9.5618  & 36.7  & 43.3 & 239.4 &  & \\
$F_5$ &  9.7788 &  9.7 & 50.8 & 28.2 & 20.7 & \\
$F_6$ &  10.0539 & 3.0 & 9.8 & 33.6 & 5.0 & \\
\hline
\end{tabular}
\label{table_T1_exp_up}
\end{table}
\end{center}

\subsection{Dark and bright states}
\label{Dark_bright}
\begin{figure}[htbp]
\begin{center}
\includegraphics[width=.60\textwidth]{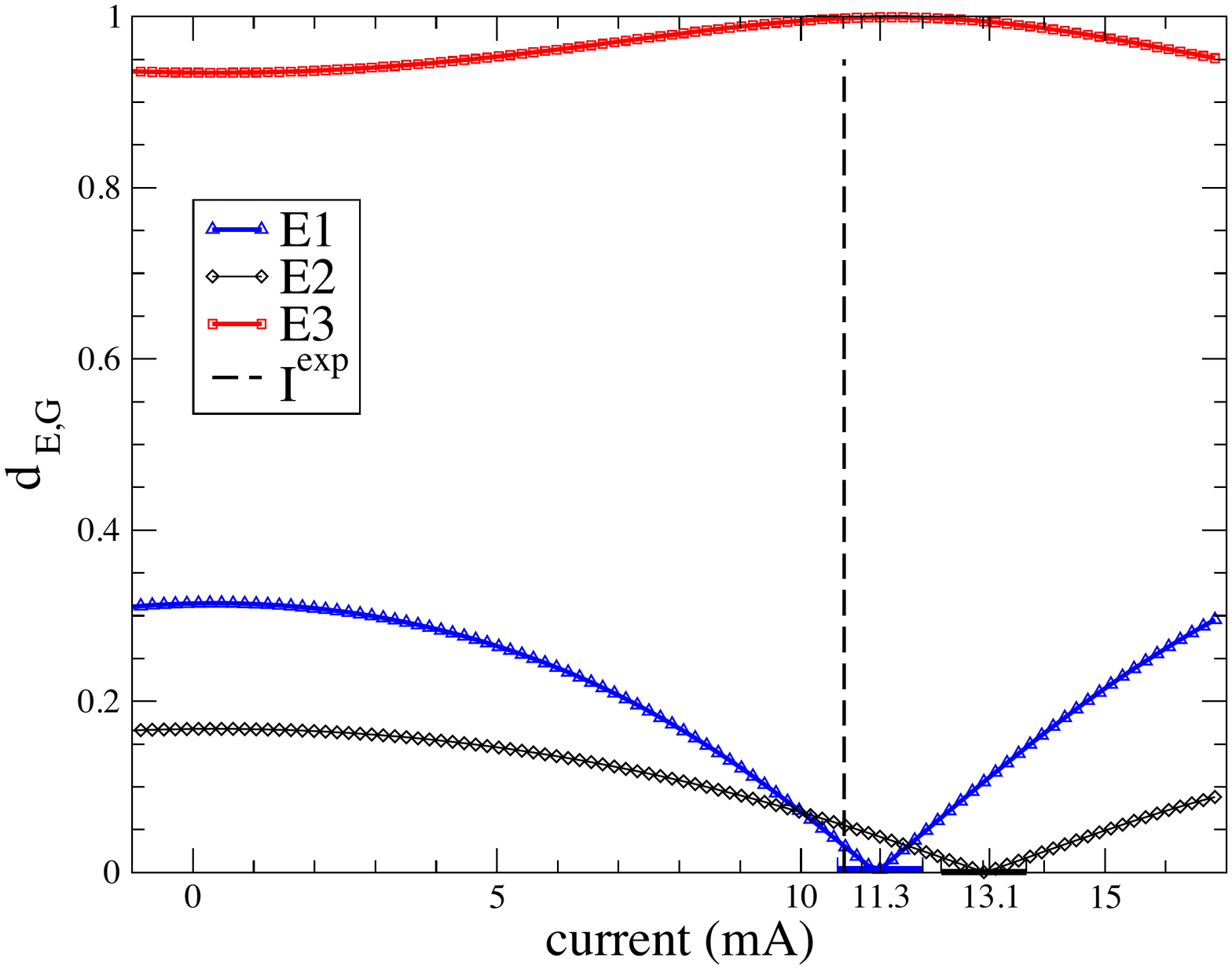}
\caption{Bright and dark features of the $E$-manifold states. The figure shows the predicted coupling  $d_{S,G}$ in Eq.~(\ref{d_SG})
between a state $S$ in the single-excitation manifold and the cavity pulse. The vertical line show the experimental value for which the states $E_1$ and $E_2$ become dark (the values are very close so cannot be distinguished). The error bars for the theoretical location of the dark point for the $E_1$ and $E_2$ states are shown  along the horizontal axis.}
\label{Edark}
\end{center}
\end{figure}

Coherently preparing an array eigenstate is accomplished via pulses applied to the same port of the cavity used to perform cooling and state measurement.  Thus, the response of the system to this drive is contained in the operator $H_{B,\mathrm{drive}}$ discussed in the previous section (see Eq.~(\ref{H_B_drive}) and its surroundings).  In the dressed basis, this operator is $H_{B, \rm{drive}}=\hbar\epsilon(t)\sum_{l=1}^L M_{0 l} (B^\dagger_{l}+B_{l})$, so the relevant value(s) of $M_{0l}$ for a particular eigenstate determine the magnitude of the response.  For the single-excitation manifold, this is an easy calculation, as the operators $B^\dagger_l$ create a single-particle eigenstate from the vacuum.  For the two-particle manifold the calculation is more involved, as the eigenbasis of $B^\dagger_l B_l$
does not coincide with the two-particle eigenmodes.  A second, equivalent method is to work in the bare basis and compute directly the matrix element of the operator $H_{\rm int}=\hbar \sum_{j=1}^3 g_j(b_j  a^\dagger +b_j^\dagger a)$ between the states $\ket{S,0_{\rm ph}}$ and $\ket{G,1_{\rm ph}}$, for desired array eigenstate $\ket{S}$.   This calculation yields
\be\label{d_SG}
d_{S,G}=|\langle \Psi_S |H_{\rm int}|\Psi_G\rangle|=\hbar \left |\langle 0_{\rm ph} |a|1_{\rm ph}\rangle\langle S|\sum_{j=1}^3 g_j b_j^\dagger |G\rangle \right|.
\ee
In Fig.~\ref{Edark} we plot this coupling $d_{E,G}$ for the $\ket{E}$-states to ground state $G$,  as a function of flux (current in the coil).  The figure shows that the theory is qualitatively in agreement with the measurements. $\ket{E_3}$  is  predicted to be always bright, in agreement with the spectroscopy image Fig~\ref{fig:SuppSpec}. Both $\ket{E_1}$ and $\ket{E_2}$ states instead become dark at a specific value of the flux (current in the coil). There is some uncertainty in the location of this dark spot due to the uncertainty $\Delta g_j=\pm 7$MHz in the measured values of the couplings $g_j$, which affects this result significantly. Theoretically we find: $I_{\rm dark }(E_1)=11.3\pm 0.7$mA and  $I_{\rm dark }(E_2)=13.1\pm 0.7$mA. These predicted values are shown, with the corresponding error bar, in Fig.~\ref{Edark}.
Experimentally the measured values are: $I^{\rm exp}_{\rm dark }(E_1)=10.71$ mA and $I^{\rm exp}_{\rm dark }(E_2)=10.64$mA; these are marked with a single dashed vertical line in Fig.~\ref{Edark} (the two lines are too close to be distinguished on the scale of the graph).
The measured value $I^{\rm exp}_{\rm dark }(E_1)$ fits within the uncertainty range of the theoretical prediction; however, $I^{\rm exp}_{\rm dark }(E_2)$ is somewhat off the theoretical prediction, and it is also slightly smaller than $I^{\rm exp}_{\rm dark }(E_1)$, contrary to what the theory would predict.

\section{Experimental Details}

\subsection{Device Parameters}
\label{Device_param}
Our device consists of three transmon qubits~\cite{PhysRevA.76.042319} on a single silicon chip.  Each qubit is formed by two aluminum paddles, connected by either a single double-angle-evaporated Al/AlO$_x$/Al Josephson junction (middle qubit) or a superconducting quantum interference device (SQUID) consisting of two junctions (outer qubits). The qubit array is located in the center of a copper waveguide cavity with dressed frequency $\omega_c/2\pi = 7.116$ GHz and $\kappa/2\pi = 10$ MHz.  Each qubit couples to the cavity via a Jaynes-Cummings $\hat{\sigma}_x(\hat{a}+\hat{a}^\dagger)$ interaction with strength $g_i$.  Because the middle qubit is located in the center of the cavity where the $\vec{E}$-field strength is greatest and its paddles are longer, it couples to the cavity more strongly than the outer qubits, with strengths $g_{\rm mid} /2\pi = $ 264 $\pm$ 7 MHz and $g_{\rm out} = /2\pi $ 155/149 $\pm$ 7. The outer qubits are characterized by a charging energy $E_c/h = 214$ MHz, which also gives us the absolute value of the anharmonicity for a transmon ($\alpha=-E_c/h$), and have a Josephson energy which gives, at zero flux, $\omega_{q1}/2\pi = $ 5.074 GHz for the left qubit and $\omega_{q3}/2\pi =$ 5.165 GHz for the right.  For the middle qubit, $E_c/h = 240$ MHz and the qubit frequency is $\omega_{q2}/2\pi= 4.892$ GHz.  The qubits are spaced by 1 mm, giving a nearest-neighbor coupling strength of $J/h = 177$ MHz and a next-nearest-neighbor coupling stength of $J_{13}/h = 26$ MHz, with uncertainties of a couple of MHz mainly due to the uncertainty in the calibrated $g_i$ values.  The qubit-cavity couplings $g_i$ and the qubit charging energy $E_c/h$ were independently calibrated by suppressing the qubit-qubit interactions. To do so we attached two coils to the cavity. One, wraped around the whole cavity, gave a uniform field. The other was fixed off center on top of the cavity, which produced a gradient field. Using the combination of these two coils we tuned the qubits such that they are effectively not interacting, $\Delta_{q_i q_j} > 10J_{q_i q_j}$. The qubit frequencies at zero flux as well as the qubit-qubit couplings were obtained by fitting the spectroscopically-measured eigenergies of the ten lowest-lying eigenstates of the array to the Bose-Hubbard Hamiltonian, as described in the next section.

\begin{figure}
\includegraphics[totalheight=0.4\textwidth]{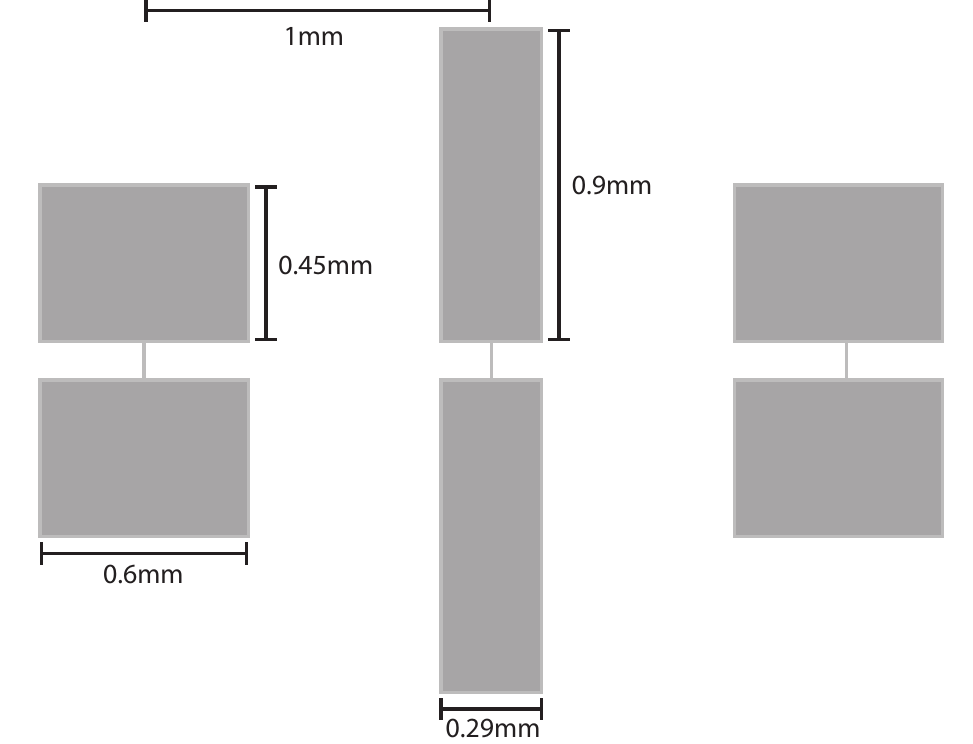}
\caption{To scale layout and dimensions of the chip with the three transmons. Josephson juctions are not illustrated}
\label{fig:Layout}
\end{figure}

The dimensions and layout of the chip placed in the cavity are shown in Fig. ~\ref{fig:Layout}.  As can be seen, the transmons have slightly different dimensions, so that they interact with the cavity with different strengths (since the field of the fundamental cavity mode is roughly uniform over the dimensions of the small chip, the difference in interaction strengths comes primarily from the different antennae configurations).  In our chip the interaction between the cavity and the transmon in the middle is nearly twice as strong as that between the cavity and the transmons on the end of the array (recall from the spectroscopy results presented in the main text that the measured $g_{2}$ was $264$ MHz while the measured $g_{1,3}$ was $150$ MHz).  Numerical simulations indicated that this mismatch in coupling strengths would improve achievable cooling rates.  

Coupling between the transmons themselves comes from two sources: cavity-mediated interactions and direct capacitve (or dipole-dipole) coupling.  Cavity-mediated interactions arise when both qubits couple to the cavity mode, as described in ref.~\cite{Majer2007}.  This coupling strength is  $J_{ij~\mathrm{cavity}}=\frac{1}{2} g_i g_j (\frac{1}{\Delta_i}+\frac{1}{\Delta_j})$, about 20 MHz for adjacent qubits and 10 MHz for edge-edge coupling, at the working point of the experiment.  Direct dipole-dipole coupling, discussed in refs.~\cite{PhysRevLett.110.030601,DalArxiv}, arises from the capacitance between transmon paddles.  In our case, this coupling is on the order of 150 MHz for adjacent qubits, and 10 MHz for the qubits on the edge. 

\subsection{Spectroscopy}

We perform spectroscopy to extract, as a function of applied magnetic flux, the eigenergies of the nine lowest-lying excited states of the array with respect to the global zero-particle ground state.  These nine states consist of three states in the single-particle manifold and six in the two-particle manifold.  As in the main text, these states are denoted $\ket{G}$, $\{\ket{E_i}\}$, and $\{\ket{F_i}\}$ respectively.  We probe the array for coil currents between -2 and +17 mA.

\begin{figure}
\includegraphics[totalheight=0.4\textwidth]{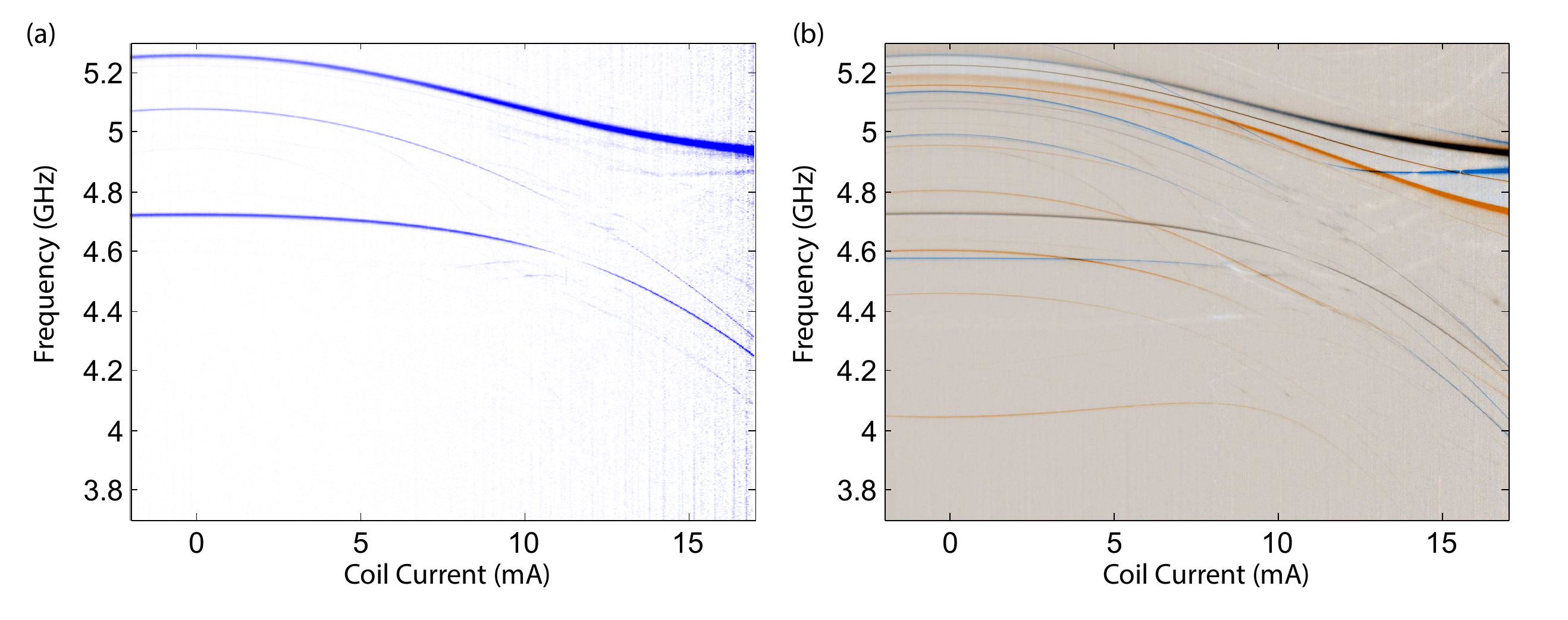}
\caption{Raw data from spectroscopy, showing (a) the $\ket{G}\rightarrow\ket{E_i}$ transitions probed with two microwave tones, and (b) the $\ket{E_1}\rightarrow\ket{F_i}$ (blue) and $\ket{E_3}\rightarrow\ket{F_i}$ (red) transitions probed with three microwave tones.}
\label{fig:SuppSpec}
\end{figure}

Since the qubit population without any excitation predominantly lies in $\ket{G}$, standard two-tone spectroscopy reveals the $\ket{G}\rightarrow\ket{E_1}$,  $\ket{G}\rightarrow\ket{E_2}$, and $\ket{G}\rightarrow\ket{E_3}$ transitions.  To perform this spectroscopy, the reflected phase of a tone near the cavity resonance (7.116 GHz) is continuously monitored as a second tone sweeps from 3.7 to 5.3 GHz.  This measurement results in Fig.~\ref{fig:SuppSpec}a, with three main lines indicating the single-particle energies.

Extraction of the two-particle energies is more involved.  For the $\ket{F_6}$ state, the energy can be directly measured via a two-photon transition from $\ket{G}$, as shown in Fig.~\ref{fig:SuppSpec}b.  For all other $\ket{F_i}$  states, however, the energies must be measured indirectly via transitions from a single-particle state.  We use $\ket{E_1}$ and $\ket{E_3}$ as stepping stones to measure the $\ket{E_1} \rightarrow \ket{F_i}$ and $\ket{E_3} \rightarrow \ket{F_i}$ transitions, by running two additional spectroscopy scans: one with the addition of a tone at the $\ket{G}\rightarrow\ket{E_1}$ frequency, and another with the addition of a tone at the $\ket{G}\rightarrow\ket{E_3}$ frequency.  The results, shown in Fig.~\ref{fig:SuppSpec}b, allow the identification of all six two-particle states.

From the extracted energies of the one- and two-particle manifolds (see Fig.~1 in the main text), we extract parameters of our device by fitting these values to predictions based on the Bose-Hubbard Hamiltonian with an additional next-nearest-neighbor coupling term. After taking into account the variation of the qubit frequencies with flux, that Hamiltonian is
\begin{equation}
\hat{H} = \hbar\sum\limits_{i = 1}^{3} \left( \omega_i(\phi) \hat{b}_i^\dagger \hat{b_i}  + \frac{\alpha_i}{2}\hat{b}^\dagger_i\hat{b}^\dagger_i\hat{b}_i\hat{b}_i \right) + \hbar J\left( \hat{b}_1^\dagger\hat{b}_2 + \hat{b}_2^\dagger\hat{b}_3 + h.c.\right) + J_{13}\left(\hat{b}_1^\dagger\hat{b}_3 + h.c. \right)
\end{equation}
with the qubit frequency as a function of flux described by $\omega(\phi) = \omega_{0i}\sqrt{\cos{\left(B_iI+A\right)}}$ for the outer qubits with SQUIDS (for the middle qubit $\omega$ is constant).  The parameters $B_i$ for each edge qubit are the ratio between the current applied to the coil and the flux threading the qubit's SQUID loop, and $A$ is an overall offset due to potential flux trapped in the SQUID loops during cooldown.

\subsection{Dispersive readout of the array}

Due to the dispersive coupling between the qubits and the cavity, each array eigenstate induces a shift in the resonant frequency of the readout cavity.  The frequency corresponding to the array in $\ket{G}$ is measurable simply via the reflected phase measurement on a network analyzer, as the array is in its global ground state without excitation pulses.  This frequency is 7.116 GHz when our system is biased up to 10 mA.  To measure the resonator frequencies corresponding to the excited states, we use microwave pulses to prepare the array in the desired eigenstate $\ket{i}$, then measure the reflected phase $\theta_i$ of a 7.116 GHz tone, referenced to the reflected phase with the array in $\ket{G}$.  This measurement was done using our LJPA in phase-preserving mode.  The standard equations for a reflected phase shift from a resonator yield that the frequency shift $\chi_i$ for a given eigenstate is related to $\theta_i$ by the equation $\chi_i = \kappa/2\tan\left(\theta_i/2\right)$.  The measured reflected phase angle $\theta_{\rm exp}$ and the corresponding $\chi_{\rm exp}$ are shown in Table \ref{table_chi}.
In the same table are also shown the $\chi$ shifts calculated theoretically according to Eqs.~(\ref{chi_E}) and (\ref{chi_F}).
Except for $\ket{F_3}$ and $\ket{F_4}$, all of the states are resolvable.  In fact, by using the LJPA in phase-sensitive mode and adjusting the measurement frequency and amplification axis (phase of the detected quadrature), we can obtain additional separation between states of interest for a particular experimental run.

To extract the population of a given state after a particular experimental sequence, we repeat the sequence several ($\sim$ 1 million) times and histogram the measured phase or quadrature amplitude values.  After the run, we take a set of calibration histograms, in which we prepare all ten states and immediately make a measurement with the same frequency and amplification axis used in the experiment.  We then fit the measured histograms to a sum of Gaussians with the same mean and variances as the calibration histograms, and from the amplitudes of each Gaussian, extract the corresponding state's population during that run.  See the next section for an example of this procedure.

\begin{center}
\begin{table}
\caption{$\chi$ shifts}
\begin{tabular}{c |c| c| c}
\hline
state & $\theta_{\rm exp}$(rad) & $\chi_{\rm exp}/\kappa$ & $\chi_{\rm th}/\kappa$   \\
\hline
E1 & 1.37 & 0.41&  0.49   \\
E2 & 0.74 &  0.19 & 0.26  \\
E3 & 1.43 & 0.43 & 0.48  \\
\hline
F1 &  2.09 & 0.86 & 1.07 \\
F2 &  1.64 & 0.53 & 0.68  \\
F3 & 1.82 & 0.64 & 0.75\\
F4 &  1.77     & 0.61 & 0.70  \\
F5 & 2.03  & 0.80 & 0.82 \\
F6 &  2.16  & 0.93 & 0.90  \\
\hline
\end{tabular}
\label{table_chi}
\end{table}
\end{center}

\begin{figure}[htbp]
\begin{center}
\includegraphics[width=.60\textwidth]{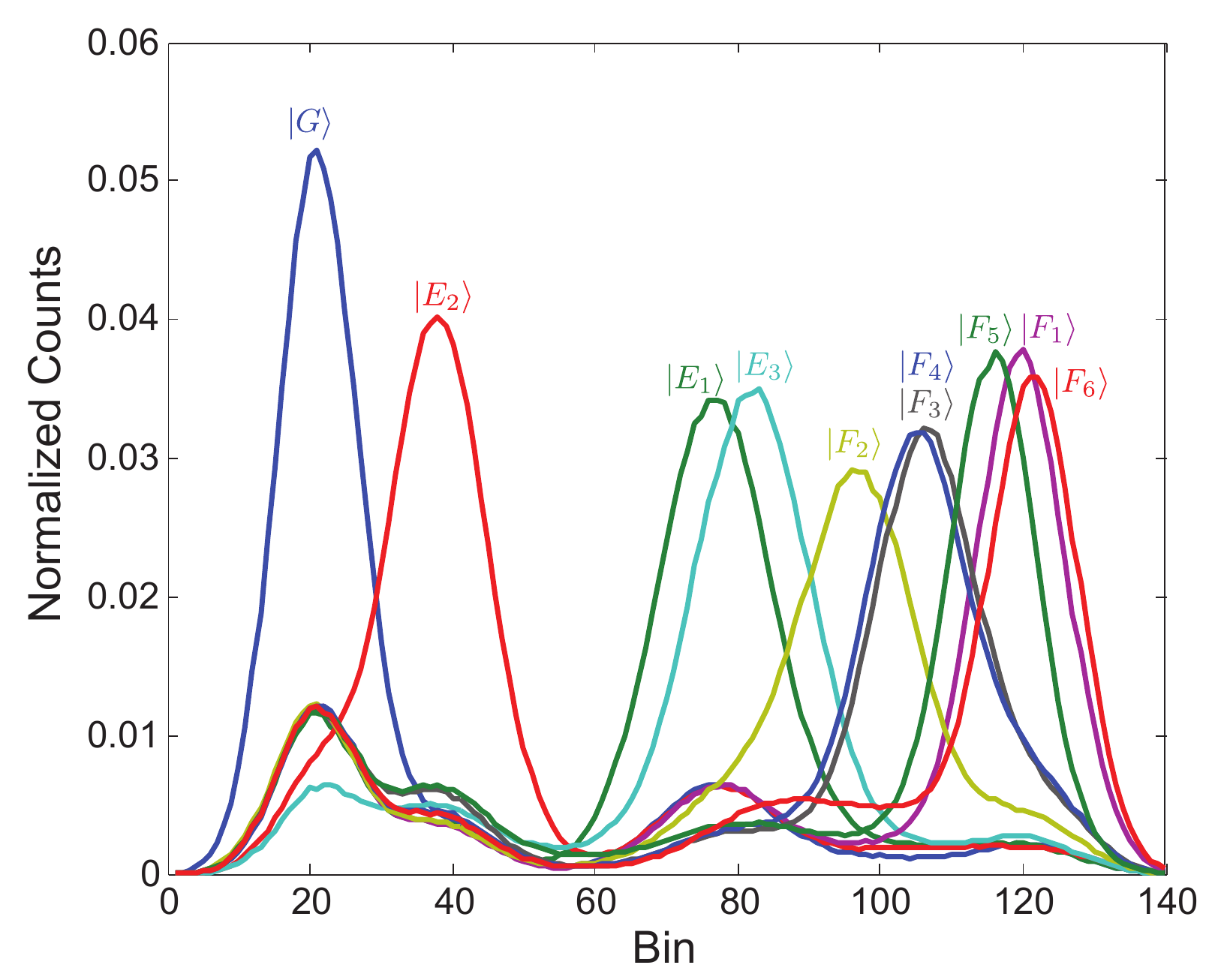}
\caption{An example of calibration histograms.}
\label{CalibHisto}
\end{center}
\end{figure}

\subsection{Extraction of Decay Rates}

Natural decay dynamics were measured by coherently initializing the system in the state of interest, waiting for a time, then measuring the array state, as described above.  We fit the population dynamics for each state to a decay model given by the matrix rate equation $\partial_t \vec{c} = \Gamma \vec{c}$, where $\vec{c}$ is a vector containing all of the state populations as a function of time and $\Gamma$ a matrix with transition rates between states.  This procedure is similar to that used by Peterer et al.~\cite{PhysRevLett.114.010501} in their study of a single transmon qubit.  This model suppresses transitions from the two-particle manifold directly to the zero-particle ground state based on the results in Ref.~\cite{PhysRevLett.114.010501}.  We also suppress transitions between states in the same manifold, since these transition frequencies are on the order of a few hundred MHz, and the photon shot noise spectrum which plays a dominant role in dissipation for this system has very little support at these frequencies. Best-fit parameters are given in \ref{table_T1_exp_down}, and were used to generate the natural decay map in Fig.~2 of the main text.  For the bath engineering decays, a similar model was used for the fit, with the addition of parameters for the intramanifold decay; all intermanifold rates were held fixed to the previously-measured natural values.

Errors in the fit, primarily at low population, occur likely due to effects such as spontaneous $T_1$ decay which cause the readout histograms to be skewed from their nominally Gaussian shape.  See \cite{GirvinT1} for a detailed explanation of this effect.  As can be seen in the main text, this effect seems to occur more often in the case of the natural decays than during the cooling protocols.  

\subsection{Calibration of Bath Engineering Drive Power}

Calibration of powers used to drive the bath-engineering transitions is done in the standard circuit-QED manner~\cite{PhysRevA.74.042318}:  first the qubit-cavity $\chi$-shift is calibrated, then with this value known, the intracavity photon number is then inferred from the Stark shift on the qubit.  To do this, in our system, we tune the edge qubits to below 3 GHz, decoupling from the middle qubit.  We then measure the $\chi$-shift between the middle qubit and the cavity by measuring, for a variety of incident powers at the cavity frequency, both the measurement-induced dephasing rate and the resulting Stark shift.  As shown in Ref.~\cite{PhysRevA.74.042318}, the measurement-induced dephasing rate is $8\chi^2\bar{n}/\kappa$, while the Stark shift is $2\chi\bar{n}$, so by comparing the slopes of these lines vs. power, we extract $\chi$ (since $\kappa$ is known).  From there, we use the Stark shift to calibrate the intracavity photon number over the range of frequencies and powers used in the bath-engineering experiment.

\subsection{Comparison of Cooling and Coherent Driving to Excite a Dark State}

Here we include a brief description of our attempts to populate dark states using coherent microwave drives rather than via the autonomous feedback protocol discussed at the end of the main paper.  

In theory, without dephasing and dissipation, and in the absence of higher Jaynes-Cummings energy states, one could drive coherent transitions to the desired state except at the singular flux bias point where the matrix element of this transition is exactly zero.  In practice, for experimentally available drive powers, driving an almost-dark transition may take such a long time that dissipation (qubit $T_1$, $T_\phi$, cavity loss) reduces the fidelity to unacceptable values.  This was the case at a flux bias of 10 mA (where the single-particle states were almost dark), where, in contrast to the usual coherent pulses which take tens of nanoseconds (we used 64 ns for the coherent pulses), the pulse took around 1 $\mu$s to excite the $|G\rangle \rightarrow  |E_1\rangle$ transition, and this caused our fidelity to be limited to about 65\%  (compare the cooling process, which achieved a fidelity for the single-excitation state of 80\%).

Another limitation of the coherent excitations is that, due to higher levels in the spectrum, off-resonant transitions which have a higher dipole moment than the dark transition may be driven before the desired transition.  This was the case at 10.71 mA (``complete'' darkness), where even at the maximum power we could excite with, we saw no population in the $\ket{E_1}$ state, but at these powers higher states (either in the two-excitation subspace or in an even higher manifold; from the dispersive shifts it was difficult to tell properly) began to get populated.

\bibliography{cooling_v2}
\bibliographystyle{ieeetr}

\end{document}